\documentclass[twocolumn]{aastex631}

\usepackage{longtable}

\usepackage[]{graphicx}
\usepackage{amsmath}
\usepackage{amssymb}
\usepackage{csquotes}
\usepackage{hyperref}

\usepackage[T1]{fontenc}

\usepackage{bm}% bold math
\usepackage{url}
\usepackage{soul}
\shorttitle{BASS LIII: Eddington ratio as the regulator of X-ray fraction in AGN}

\shortauthors{Gupta et al.}

\begin{document}

\title{BASS LIII: The Eddington Ratio as the Primary Regulator of the Fraction of X-ray Emission in Active Galactic Nuclei}

\author[0009-0007-9018-1077]{Kriti Kamal Gupta}
\thanks{E-mail: kkgupta@uliege.be}
\affiliation{Instituto de Estudios Astrof\'isicos, Facultad de Ingenier\'ia y Ciencias, Universidad Diego Portales, Av. Ej\'ercito Libertador 441, Santiago, Chile}
\affiliation{STAR Institute, Li\`ege Universit\'e, Quartier Agora - All\'ee du six Ao\^ut, 19c B-4000 Li\`ege, Belgium}
\affiliation{Sterrenkundig Observatorium, Universiteit Gent, Krijgslaan 281 S9, B-9000 Gent, Belgium}

\author[0000-0001-5231-2645]{Claudio Ricci}
\affiliation{Instituto de Estudios Astrof\'isicos, Facultad de Ingenier\'ia y Ciencias, Universidad Diego Portales, Av. Ej\'ercito Libertador 441, Santiago, Chile} 
\affiliation{Kavli Institute for Astronomy and Astrophysics, Peking University, Beijing 100871, China}

\author[0000-0003-3450-6483]{Alessia Tortosa}
\affiliation{INAF\textemdash Osservatorio Astronomico di Roma, via di Frascati 33, I-00078 Monte Porzio Catone, Italy}

\author[0000-0001-8433-550X]{Matthew J. Temple}
\affiliation{Centre for Extragalactic Astronomy, Department of Physics, Durham University, South Road, Durham DH1 3LE, UK}

\author[0000-0001-8433-550X]{Michael J. Koss}
\affiliation{Eureka Scientific, 2452 Delmer Street, Suite 100, Oakland, CA 94602-3017, USA}

\author[0000-0002-3683-7297]{Benny Trakhtenbrot}
\affiliation{School of Physics and Astronomy, Tel Aviv University, Tel Aviv 69978, Israel}
\affiliation{Max-Planck-Institut f{\"u}r extraterrestrische Physik, Gie\ss{}enbachstra\ss{}e 1, 85748 Garching, Germany}
\affiliation{Excellence Cluster ORIGINS, Boltzmannsstra\ss{}e 2, 85748, Garching, Germany}

\author[0000-0002-8686-8737]{Franz E. Bauer}
\affiliation{Instituto de Alta Investigaci\'on, Universidad de Tarapac\'a, Casilla 7D, Arica, Chile}

\author[0000-0001-7568-6412]{Ezequiel Treister}
\affiliation{Instituto de Alta Investigaci\'on, Universidad de Tarapac\'a, Casilla 7D, Arica, Chile}

\author[0000-0002-7962-5446]{Richard Mushotzky}
\affiliation{Department of Astronomy and Joint Space-Science Institute, University of Maryland, College Park, MD 20742, USA}

\author[0000-0002-0273-218X]{Elias Kammoun}
\affiliation{Cahill Center for Astrophysics, California Institute of Technology, 1216 East California Boulevard, Pasadena, CA 91125, USA}

\author[0000-0001-6264-140X]{Iossif Papadakis}
\affiliation{Department of Physics and Institute of Theoretical and Computational Physics, University of Crete, 71003 Heraklion, Crete, Greece}
\affiliation{Institute of Astrophysics, FORTH, GR-71110 Heraklion, Crete, Greece}

\author[0000-0002-5037-951X]{Kyuseok Oh}
\affiliation{Korea Astronomy and Space Science Institute, Daedeokdae-ro 776, Yuseong-gu, Daejeon 34055, Republic of Korea}

\author[0000-0003-0006-8681]{Alejandra Rojas}
\affiliation{Departamento de F\'isica, Universidad T\'ecnica Federico Santa Mar\'ia, Vicu\~{n}a Mackenna 3939, San Joaqu\'in, Santiago, Chile}

\author[0000-0001-9910-3234]{Chin-Shin Chang}
\affiliation{Observatoire de Gen\`eve, Universit\'e de Gen\`eve, Chemin Pegasi 51, 1290, Versoix, Switzerland}

\author[0000-0002-8604-1158]{Yaherlyn Diaz}
\affiliation{Instituto de Estudios Astrof\'isicos, Facultad de Ingenier\'ia y Ciencias, Universidad Diego Portales, Av. Ej\'ercito Libertador 441, Santiago, Chile}

\author[0000-0001-7500-5752]{Arghajit Jana}
\affiliation{Instituto de Estudios Astrof\'isicos, Facultad de Ingenier\'ia y Ciencias, Universidad Diego Portales, Av. Ej\'ercito Libertador 441, Santiago, Chile}

\author[0000-0002-2603-2639]{Darshan Kakkad}
\affiliation{Centre for Astrophysics Research, University of Hertfordshire, Hatfield, AL10 9AB, UK}

\author[0000-0001-8931-1152]{Ignacio del Moral-Castro}
\affiliation{Instituto de Astrof\'isica, Facultad de F\'isica, Pontificia Universidad Cat\'olica de Chile, Av. Vicu\~na Mackenna 4860, Santiago, Chile}

\author[0000-0003-2196-3298]{Alessandro Peca}
\affiliation{Eureka Scientific, 2452 Delmer Street, Suite 100, Oakland, CA 94602-3017, USA}
\affiliation{Department of Physics, Yale University, P.O. Box 208120, New Haven, CT 06520, USA}

\author[0000-0003-2284-8603]{Meredith C. Powell}
\affiliation{Leibniz-Institut f\"ur Astrophysik Potsdam (AIP), An der Sternwarte 16, 14482 Potsdam, Germany}

\author[0000-0003-2686-9241]{Daniel Stern}
\affiliation{Jet Propulsion Laboratory, California Institute of Technology, 4800 Oak Grove Drive, MS 169-224, Pasadena, CA 91109, USA}

\author[0000-0002-0745-9792]{C. Megan Urry}
\affiliation{Department of Physics, Yale University, P.O. Box 208120, New Haven, CT 06520, USA}
\affiliation{Yale Center for Astronomy \& Astrophysics, Yale University, P.O. Box 208120, New Haven, CT 06520-8120, USA}

\author[0000-0002-4226-8959]{Fiona Harrison}
\affiliation{Cahill Center for Astrophysics, California Institute of Technology, 1216 East California Boulevard, Pasadena, CA 91125, USA}

\date{\today} 

%------------------------------------------------------------------%
%------------------------------------------------------------------%

\begin{abstract}
Active galactic nuclei (AGN) emit radiation via accretion across the entire energy spectrum. While the standard disk and corona model can somewhat describe this emission, it fails to predict specific features such as the soft X-ray excess, the short-term optical/UV variability, and the observed UV/X-ray correlation in AGN. In this context, the fraction of AGN emission in different bands (i.e., bolometric corrections) can be useful to better understand the accretion physics of AGN. Past studies have shown that the X-ray bolometric corrections are strongly dependent on the physical properties of AGN, such as their luminosities and Eddington ratios. However, since these two parameters depend on each other, it has been unclear which is the main driver of the X-ray bolometric corrections. We present here results from a large study of hard X-ray-selected (14--195\,keV) nearby ($z<0.1$) AGN. Based on our systematic analysis of the simultaneous optical-to-X-ray spectral energy distributions of 236 unobscured AGN, we found that the primary parameter controlling the X-ray bolometric corrections is the Eddington ratio. Our results show that while the X-ray bolometric correction increases with the bolometric luminosity for sources with intermediate Eddington ratios ($0.01-1$), this dependence vanishes for sources with lower Eddington ratios ($<0.01$). This could be used as evidence for a change in the accretion physics of AGN at low Eddington ratios. 
\end{abstract}

%\maketitle

%------------------------------------------------------------------%
%------------------------------------------------------------------%
  
\section{Introduction}\label{sec:intro}

Active galactic nuclei (AGN) grow by actively accreting material onto their central supermassive black holes (SMBHs) and, in the process, emit radiation, mainly at optical/UV and X-ray energies \citep{1964ApJ...140..796S,1969Natur.223..690L,1984ARA&A..22..471R}. The optical/UV emission is attributed to the accretion flow around the central black hole, generally described by a standard, geometrically thin yet optically thick \cite{1973A&A....24..337S} disk with a temperature gradient (with the innermost regions being the hottest; see also \citealp{1973blho.conf..343N}). Higher energy X-ray radiation, on the other hand, is thought to originate in a compact corona of hot ($\sim 10 - 100\,{\rm keV}$) electrons, located close to the central SMBH. This coronal plasma up-scatters the optical/UV seed photons from the accretion disk via Comptonization to produce X-rays  \citep{1991ApJ...380L..51H}.

Although these two physical components successfully reproduce the optical-to-X-ray spectral energy distributions (SEDs) of AGN on a basic level, they fail to describe the wide diversity of AGN and certain features observed in AGN SEDs (see \citealp{1999PASP..111....1K} for a detailed overview). The X-ray spectra of unabsorbed AGN often show emission above the power law continuum below $\sim$1--2\,keV, referred to as the soft X-ray excess (e.g., \citealp{1987ApJ...314..699B}). A secondary \enquote*{warm} ($\sim$0.1--1\,keV) Comptonization region has been proposed to account for the soft excess (e.g., \citealp{1998MNRAS.301..179M,2006MNRAS.365.1067C,2012MNRAS.420.1848D,2018A&A...611A..59P,2018MNRAS.480.1247K}). Alternatively, this soft excess can also be explained by X-ray reflection from the inner regions of the accretion disk (e.g., \citealp{2006MNRAS.365.1067C}). However, these two scenarios (i.e., warm corona and reflection) are not distinguishable and could possibly coexist (e.g., \citealp{2019ApJ...871...88G}). The multi-temperature disk model is also unable to reproduce the short-term stochastic variability observed in the optical/UV for many AGN (e.g., \citealp{2001ApJ...555..775C,2018NatAs...2..102L}), as well as the strong variations in bolometric luminosity of some AGN on timescales of months to years (e.g., \citealp{2015ApJ...800..144L,2018ApJ...854..160R,2020MNRAS.491.4925G}). Furthermore, the disappearance of the UV excess, commonly called the \enquote*{big blue bump}, in low-luminosity AGN (LLAGN) is also beyond the predictions of the standard disk (e.g., \citealp{2008ARA&A..46..475H}).

Numerous studies have shown that the monochromatic X-ray and UV luminosities of AGN display a tight nonlinear correlation (e.g., \citealp{1986ApJ...305...83A,1994ApJS...92...53W,2006AJ....131.2826S,2016ApJ...819..154L}). However, the physical connection between the optical/UV disk and the X-ray coronal emission driving this correlation is still not understood (e.g., \citealp{2017A&A...602A..79L}). Over the years, multiple accretion flow and disk-corona models have been proposed to explain the possible origin of some or all of these features. A few examples include radiatively inefficient hot accretion flows (e.g., \citealp{2014ARA&A..52..529Y}), hot and warm corona model (e.g., \citealp{2018MNRAS.480.1247K}), X-ray illumination of the accretion disk (e.g., \citealp{2022ApJ...935...93P,2025A&A...697A..55K}), and magnetically dominated accretion disks (e.g., \citealp{2024OJAp....7E..20H}). While these models are able to explain the multiband emission and observed variability of some specific AGN sources, their applicability to the larger AGN population is debated.

In the last few decades, many studies have discussed the importance of bolometric corrections to better understand the AGN disk-corona interplay with respect to the total bolometric emission. By definition, bolometric corrections ($\kappa_{\lambda}$) are the ratio of the total bolometric luminosity ($L_{\rm bol}$) of an AGN to the luminosity in a specific wavelength (or energy) band ($L_{\lambda}$; $\kappa_{\lambda}=L_{\rm bol}/L_{\lambda}$). Hence, they indicate what fraction of the AGN's total energy is emitted in a specific energy band. Investigating the dependence of these bolometric corrections on the properties of AGN can help shed light on the physical processes involved in the multiwavelength emission of AGN and their evolution with the accretion rate. Past studies have discovered that, while the optical bolometric corrections do not show any significant change with properties like the black hole mass ($M_{\rm BH}$), bolometric luminosity ($L_{\rm bol}$), or the Eddington ratio ($\lambda_{\rm Edd}$) of the AGN, the X-ray bolometric corrections are strongly dependent on the latter two properties (e.g., \citealp{2009MNRAS.392.1124V,2012MNRAS.425..623L,2019MNRAS.488.5185N,2020A&A...636A..73D}). Indeed, the 2--10\,keV bolometric correction ($\kappa_{2-10}$) shows a significant increase with both $L_{\rm bol}$ and $\lambda_{\rm Edd}$. However, since these two parameters are intrinsically linked to each other ($\lambda_{\rm Edd} = L_{\rm bol}/L_{\rm Edd}$; $L_{\rm Edd} = 1.5\times10^{38} \times \frac{M_{\rm BH}}{M_{\odot}}\,\rm erg\,s^{-1}$) it has been challenging to disentangle them and constrain their individual effects on bolometric corrections. Therefore, we still lack a clear understanding of the main physical property that controls the bolometric corrections and regulates the accretion and emission physics of AGN. 

\defcitealias{2024A&A...691A.203G}{G24}

The main aim of this work is to study the effect of the physical parameters of AGN (i.e., $M_{\rm BH}$, $L_{\rm bol}$, and $\lambda_{\rm Edd}$) on the 2--10\,keV X-ray bolometric corrections to determine the primary regulator of the fraction of X-ray emission in AGN. To do so, we used the largest sample of hard X-ray-selected, nearby, unobscured AGN, with simultaneous optical, UV, and X-ray data, described in detail in Sect. \ref{sec:data}. We present our main results in Sect. \ref{sec:bc} and discuss their implications in the context of AGN accretion in Sect. \ref{sec:discuss}. Finally, we summarize our findings in Sect. \ref{sec:summary}. Throughout the paper, we assume a cosmological model with $H_{\rm 0} = 70\,\rm km\,s^{-1}\,Mpc^{-1}$, $\Omega_{\rm M} = 0.3$, and $\Omega_{\lambda} = 0.7$. All correlations were obtained using various functions from the \texttt{statistics}\footnote{\url{https://docs.scipy.org/doc/scipy/reference/stats.html}} module of the Python library \texttt{scipy} \citep{2020SciPy-NMeth} and the Python package \texttt{linmix}\footnote{\url{https://linmix.readthedocs.io/en/latest/}} \citep{2007ApJ...665.1489K}. Specifically, the significance of the correlations is determined using the Pearson's correlation test.

%------------------------------------------------------------------%
%------------------------------------------------------------------%

\section{Sample and data}\label{sec:data}

Our sample is based on the 70-month Swift/Burst Alert Telescope (BAT) catalog \citep{2013ApJS..207...19B} consisting of 858 AGN. From this, we selected a sample of 236 unobscured AGN in the local Universe ($z<0.1$), based on their line-of-sight column densities ($N_{\rm H} < 10^{22}\,{\rm cm^{-2}}$; \citealp{2015ApJ...815L..13R,2017ApJS..233...17R}). The 14--195\,keV hard-X-ray detector of Swift/BAT \citep{2005SSRv..120..143B,2013ApJS..209...14K} is ideal to detect optically-faint AGN with strong X-ray emission, which might be missed by optical surveys. Importantly, all the AGN detected by Swift/BAT have been observed simultaneously in the soft X-rays (0.3--10\,keV energy range) and optical/UV (six filters spanning 170 to 650 nm) by the X-ray telescope \citep{2005SSRv..120..165B} and the UV/optical telescope \citep{2008MNRAS.383..627P,2010MNRAS.406.1687B} on board Swift, thus providing contemporaneous multiband data that are exceptionally well suited for broadband SED analysis.

In a previous study (\citealp{2024A&A...691A.203G}; hereafter \citetalias{2024A&A...691A.203G}), we performed a comprehensive analysis of the optical-to-X-ray SEDs for our AGN sample. Here we briefly summarize the key aspects of the SED construction and modeling, but for a detailed discussion, please refer to \citetalias{2024A&A...691A.203G}. We created the multiband SEDs by combining the 0.3--10\,keV X-ray spectrum (extracted following the Swift/XRT pipeline) and the optical/UV fluxes (in at least four Swift/UVOT bands) of our sources. The latter were corrected for contamination from the AGN host galaxy following a careful image decomposition procedure using \textsc{GALFIT} \citep{2002AJ....124..266P,2010AJ....139.2097P}. The final set of SEDs was fitted using a combination of models to describe the X-ray coronal and the optical/UV disk emission. The X-ray part of the SED was fitted using an absorbed power law with a reflection component. A blackbody component that accounts for the soft excess, or a partially covering warm absorber, was added if and when needed. The lower energy side of the SED was modeled with a dust-extincted multi-temperature accretion disk model with a fixed inner radius of $6R_{\rm g}$. The bolometric luminosities for all AGN were then calculated by adding the intrinsic X-ray luminosity in the 0.1--500\,keV range and the intrinsic optical/UV luminosity in the $1000\mu \rm m$ to 0.1\,keV range. The infrared emission was excluded because it is driven by reprocessed UV and optical radiation. We also calculated the intrinsic luminosity in the 2--10\,keV X-ray band to estimate 2--10\,keV X-ray bolometric corrections ($\kappa_{2-10}=L_{\rm bol}/L_{2-10}$). Black hole mass measurements obtained using either the virial method or direct reverberation mapping are available for 234/236 AGN thanks to the BAT AGN Spectroscopic Survey\footnote{\url{www.bass-survey.com}} \citep{2017ApJ...850...74K,2022ApJS..261....2K,2022ApJS..261....5M}. We used these estimates to calculate the Eddington ratios for our AGN sample.

Following the details mentioned above, our AGN sample spans five orders of magnitude in both bolometric luminosity and Eddington ratio, specifically including a sufficient number of sources (36/236) with low luminosities ($L_{\rm bol} < 10^{44}\,\rm erg\,s^{-1}$) and low Eddington ratios ($\lambda_{\rm Edd} < 0.01$) to study if and how their accretion properties change with respect to the rest of the sample.

%------------------------------------------------------------------%
%------------------------------------------------------------------%

\begin{figure*}
    %\begin{subfigure}[t]{0.49\textwidth}
    %\centering
    \includegraphics[width=0.49\textwidth]{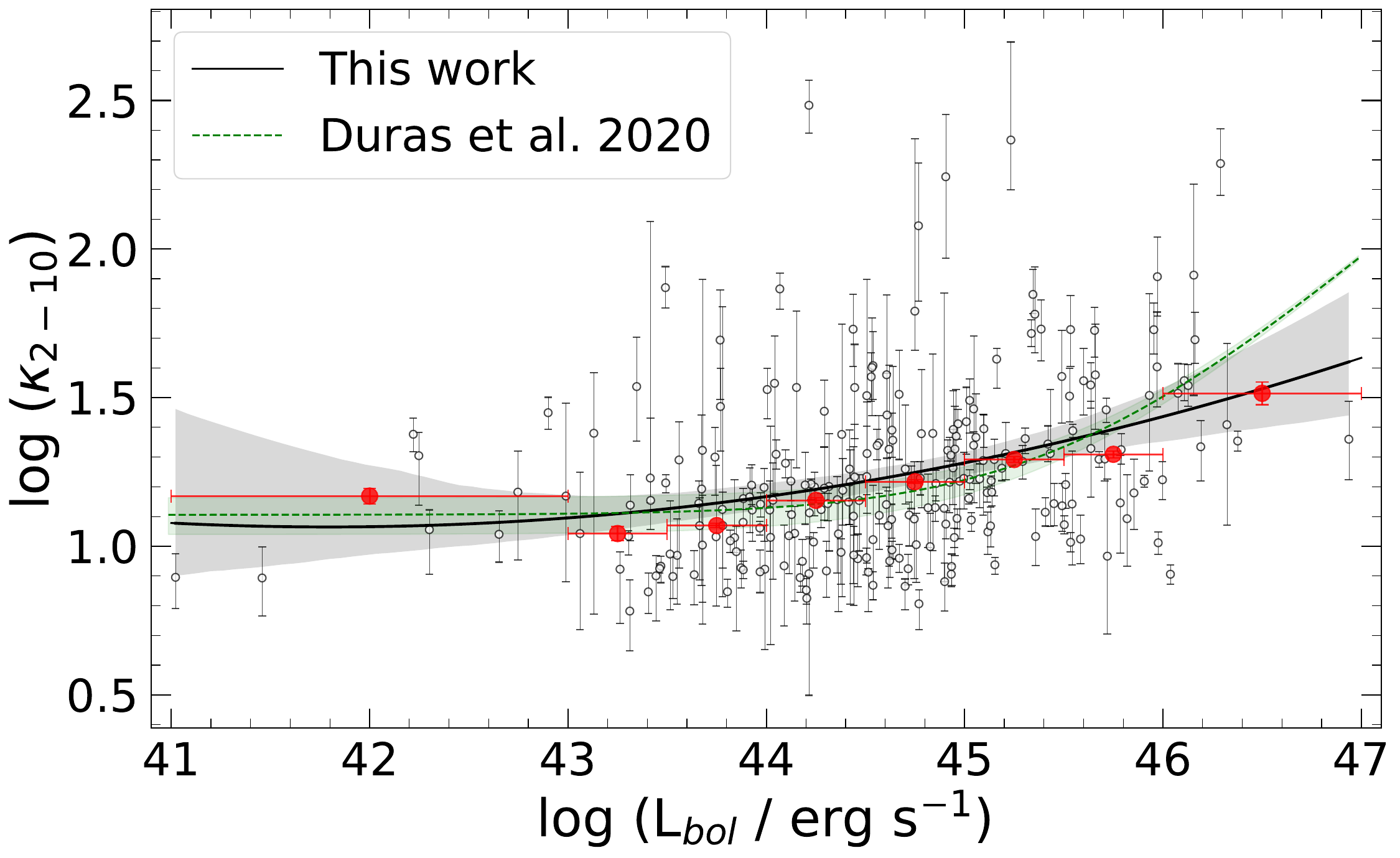}
    %\caption{}
    \label{fig:kx_lb}
    %\end{subfigure}
%\hspace{-1mm}
    %\begin{subfigure}[t]{0.49\textwidth}
    %\centering
    \includegraphics[width=0.49\textwidth]{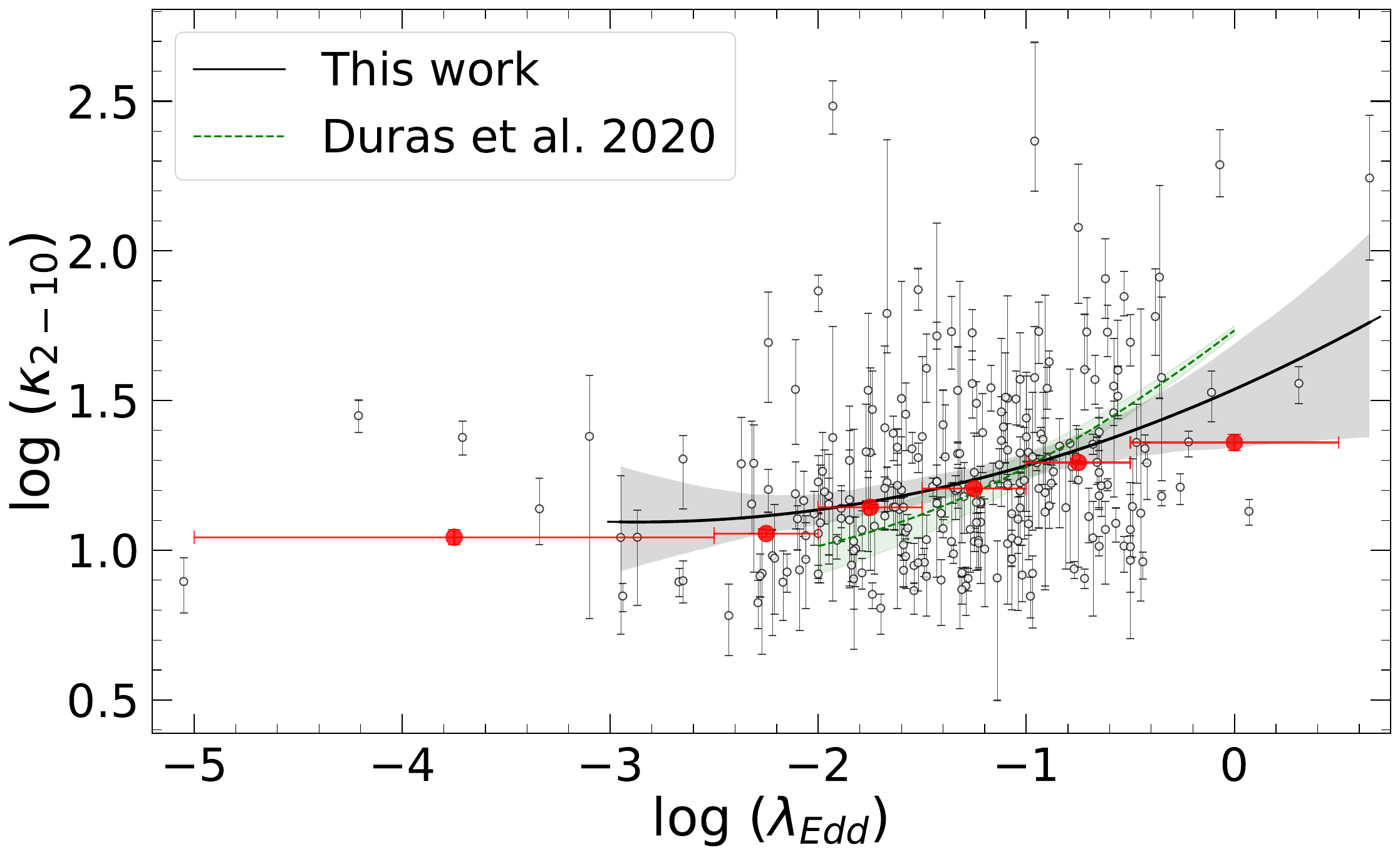}
    %\caption{}
    \label{fig:kx_er}
    %\end{subfigure}
\vspace{5mm}
    %\begin{subfigure}[t]{\textwidth}
    \centering
    \includegraphics[width=0.5\textwidth]{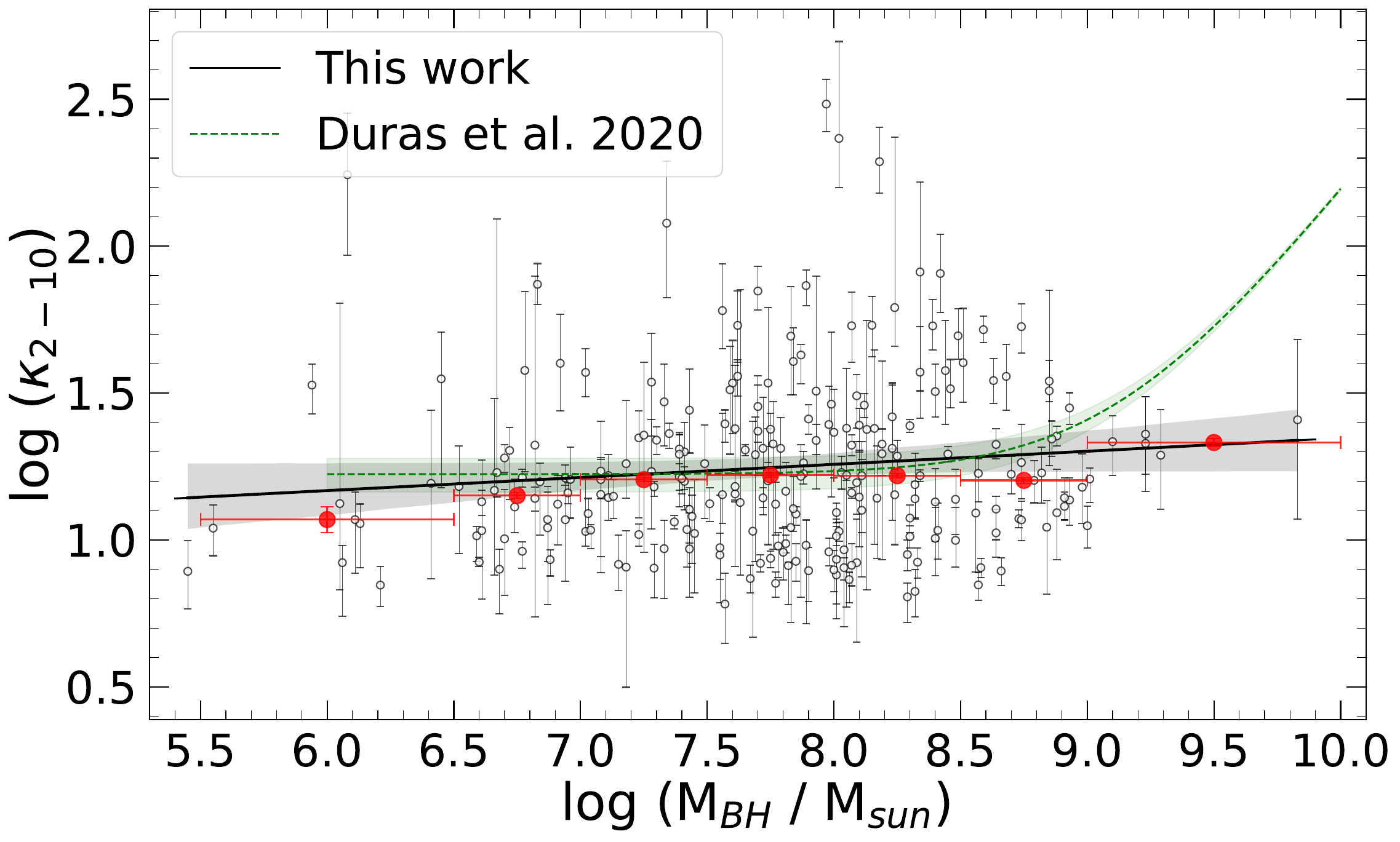}
    %\caption{}
    \label{fig:kx_mbh}
    %\end{subfigure}
    
\caption{Relation between the 2--10\,keV bolometric correction and AGN properties. We show $\kappa_{2-10}$ as a function of the bolometric luminosity (top left panel), the Eddington ratio (top right panel), and black hole mass (bottom panel) for our sample of hard-X-ray-selected unobscured AGN. We found an increase in $\kappa_{2-10}$ with $L_{\rm bol}$ and $\lambda_{\rm Edd}$, but did not find any dependence of $\kappa_{2-10}$ on $M_{\rm BH}$. The solid black line shows the best-fit relation, while the shaded gray region marks the one-sigma confidence interval. We also compare with previous results \citep{2020A&A...636A..73D} as the dashed green line. The red points are the median $\kappa_{2-10}$ values in bins of $L_{\rm bol}$, $\lambda_{\rm Edd}$, and $M_{\rm BH}$, respectively, such that each bin has a minimum of six sources.}
\label{fig:kx}
\end{figure*}

%------------------------------------------------------------------%
%------------------------------------------------------------------%
 
\section{The primary regulator of the X-ray bolometric correction}\label{sec:bc}

In this section, we discuss the dependence of the X-ray bolometric correction ($\kappa_{2-10}$) on different physical properties of AGN, such as bolometric luminosity (Sect. \ref{sec:bc_vs_lbol}), Eddington ratio (Sect. \ref{sec:bc_vs_er}), and black hole mass (Sect. \ref{sec:bc_vs_mbh}). In Sect. \ref{sec:bc_bin}, we attempt to disentangle the effects of these three parameters to determine the main driver of $\kappa_{2-10}$ in AGN.  We have also compared our results with \citet{2020A&A...636A..73D}, who performed a similar broadband SED study of $\sim 1000$ AGN to estimate their multiband bolometric corrections and investigate their dependence on AGN properties. 

%------------------------------------------------------------------%
%------------------------------------------------------------------%

\subsection{$\kappa_{2-10}$ vs $L_{\rm bol}$}\label{sec:bc_vs_lbol}

In Fig. \ref{fig:kx} (top left panel), we show the dependence of $\kappa_{2-10}$ on $L_{\rm bol}$. The black points represent the data from our work, while the red points show the median of $\kappa_{2-10}$ in bolometric luminosity bins, with a minimum of six sources per bin. We also show the best fit to the data as the black solid line, while the shaded gray region is the one sigma confidence interval. In agreement with previous studies (e.g., \citealp{2004IAUS..222...49M,2009MNRAS.392.1124V,2020A&A...636A..73D}), $\kappa_{2-10}$ shows a clear positive correlation with $L_{\rm bol}$ ($p$-value = $2.52 \times 10^{-7}$). This trend is best described with a second-degree polynomial fit:

\begin{equation}\label{eq:kx_lb}
    \begin{aligned}
        {\rm log}\,(\kappa_{2-10}) = &\,\,(37.75 \pm 26.15) + (-78.99 \pm 53.03) \times \mathcal{L}_{\rm bol, 45}\\
        & + (42.52 \pm 26.88) \times (\mathcal{L}_{\rm bol, 45})^2,
    \end{aligned}
\end{equation}
where, $\mathcal{L}_{\rm bol, 45} \equiv \log\,(L_{\rm bol} / 10^{45}\,\rm erg\,s^{-1})$. This fit is valid within the range of $L_{\rm bol} = 10^{41-46}\,{\rm erg\,s^{-1}}$. We compare our best-fit relation with the one from \citet{2020A&A...636A..73D}, shown as the dashed green line in Fig. \ref{fig:kx}. While the two relations agree well up to $L_{\rm bol} \sim 10^{46}\,\rm erg\,s^{-1}$, the slight deviation above those values could result from the small number of sources in our sample in that last bin (see Sect. \ref{sec:high_er} for more details). Ignoring the first and the last bins (with only a few sources) does not have any significant effect on the best-fit relation quoted above.

This strong correlation between $\kappa_{2-10}$ and $L_{\rm bol}$ and the absence of any relation between $\kappa_{2-10}$ and $L_{2-10}$ (see Sect. 7.4.1 and Fig. 20 of \citetalias{2024A&A...691A.203G}), suggests that a higher bolometric luminosity does not necessarily imply a stronger contribution from the X-rays. Conversely, sources with large accretion luminosities seem to emit less X-ray emission in the 2--10\,keV energy band, leading to an overall increase in $\kappa_{2-10}$. This also puts in perspective the anti-correlation obtained between UV bolometric corrections and $L_{\rm bol}$ (see Fig. 24 of \citetalias{2024A&A...691A.203G}). As discussed in \citetalias{2024A&A...691A.203G}, a statistically significant ($p$-value = $5.80 \times 10^{-4}$) negative dependence is observed between the UV bolometric correction in the UVW2 band ($\kappa_{\rm W2}$) and $L_{\rm bol}$, suggesting an increase in the UV luminosity as the bolometric luminosity increases. Combining both results, we can assess that as the bolometric luminosity increases, the contribution from the UV to $L_{\rm bol}$ also increases. In contrast, the X-rays do not show similar behavior. Instead, the X-rays may saturate at some point while the UV emission drives the bolometric luminosity in the brightest AGN ($L_{\rm bol} > 10^{44.5-45}\,{\rm erg\,s^{-1}}$).

%------------------------------------------------------------------%
%------------------------------------------------------------------%

\subsection{$\kappa_{2-10}$ vs $\lambda_{\rm Edd}$}\label{sec:bc_vs_er}

The positive correlation between $\kappa_{2-10}$ and $\lambda_{\rm Edd}$ is shown in Fig. \ref{fig:kx} (top right panel). The black line shows the best-fit relation to the data. We report the following second-degree polynomial equation that describes the statistically significant ($p$-value = $1.08\times10^{-7}$) dependence of $\kappa_{2-10}$ on $\lambda_{\rm Edd}$:

\begin{equation}\label{eq:kx_er}
    \begin{aligned}
        {\rm log}\,(\kappa_{2-10}) = &\,\,(1.54 \pm 0.06) + (0.31 \pm 0.09) \times {\rm log}\,(\lambda_{\rm Edd})\\
        & +(0.05 \pm 0.03) \times ({\rm log}\,[\lambda_{\rm Edd}])^2
    \end{aligned}
\end{equation}
The range of validity for the above relation is ${\rm log}\,\lambda_{\rm Edd} = -3$ to 0. $\kappa_{2-10}$ shows a flat behavior for the lower Eddington ratio bin ($-5 < {\rm log}\,\lambda_{\rm Edd} < -3$). However, since we only have have a few sources in this bin, it is really hard to comment on how $\kappa_{2-10}$ evolves in this regime. For comparison, we show the best-fit relation obtained by \citeauthor{2020A&A...636A..73D} (\citeyear{2020A&A...636A..73D}; dashed green line) along with our results (solid black line) in Fig. \ref{fig:kx}. Our correlation is relatively less steep as we have fewer sources in the highest $\lambda_{\rm Edd}$ bin (${\rm log}\,\lambda_{\rm Edd}>0$), compared to the sample of \citet{2020A&A...636A..73D}. Otherwise, the two correlations are consistent within uncertainties for sources with $\lambda_{\rm Edd}$ between 0.01 and 1. We briefly explore the behavior of $\kappa_{2-10}$ at higher Eddington ratios ($\lambda_{\rm Edd}>1$) in Sect. \ref{sec:high_er}.

%------------------------------------------------------------------%
%------------------------------------------------------------------%

\subsection{$\kappa_{2-10}$ vs $M_{\rm BH}$}\label{sec:bc_vs_mbh}

We also check if $\kappa_{2-10}$ shows any dependence on the black hole mass of the AGN in our sample (Fig. \ref{fig:kx}, bottom panel), but we do not find any significant correlation ($p$-value = 0.08). \citet{2020A&A...636A..73D} reported an increase in $\kappa_{2-10}$ with $M_{\rm BH}$, however, the correlation they found was mainly driven by the highest $M_{\rm BH}$ bin ($10^{9-10} M_\odot$). Since we only have six sources in our sample with $M_{\rm BH} > 10^9 M_\odot$, we cannot make any statistically conclusive statements regarding the dependence of $\kappa_{2-10}$ on $M_{\rm BH}$ in that regime. Over the rest of the parameter space, our results are consistent with \cite{2020A&A...636A..73D}.

%------------------------------------------------------------------%
%------------------------------------------------------------------%

\begin{figure*}
  %\begin{subfigure}[t]{\textwidth}
    \centering
    \includegraphics[width=0.6\textwidth]{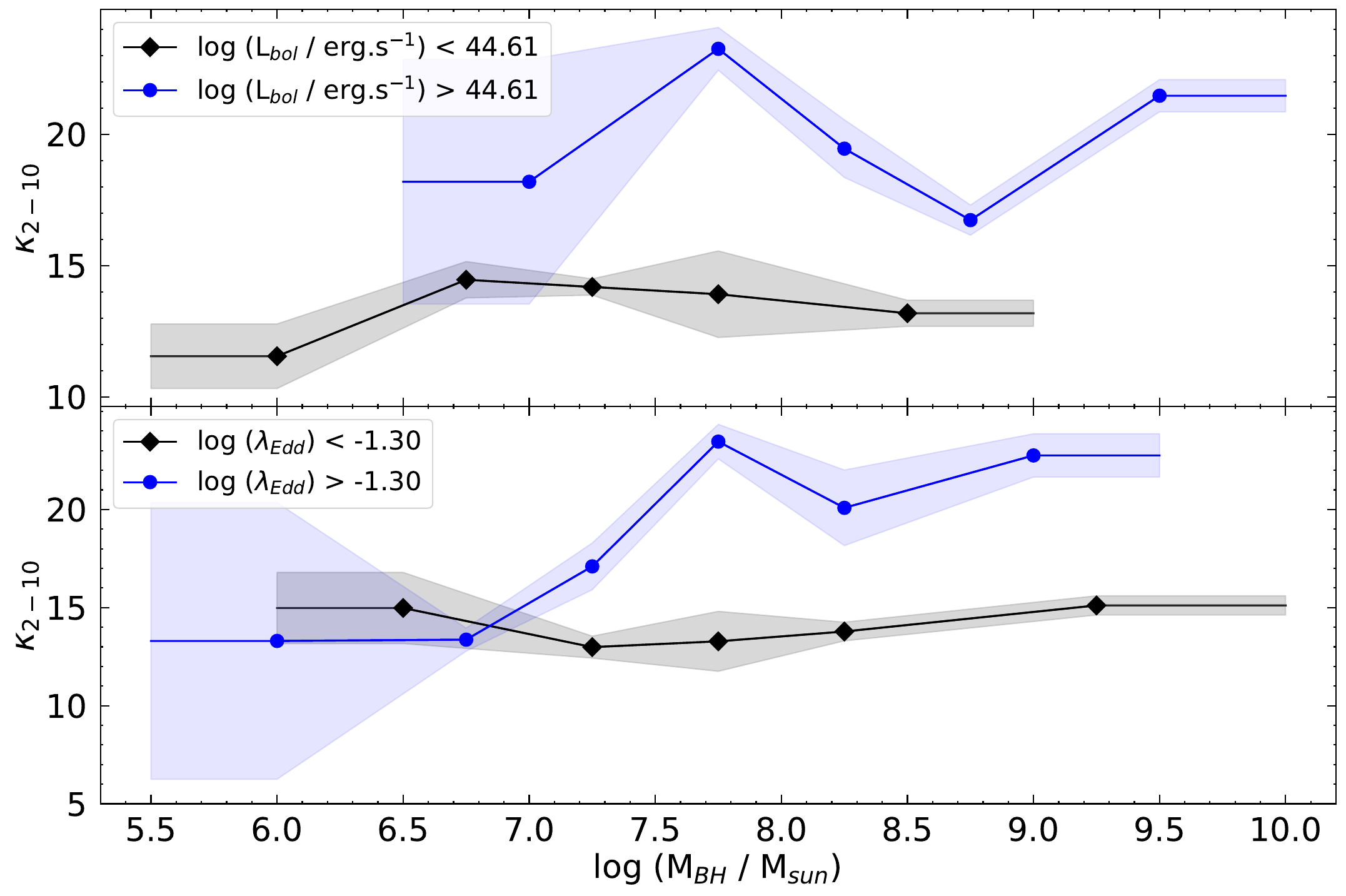} 
    %\vspace{-15mm}
    %\caption{}        
    \label{fig:kx_mbh_bin}
  %\end{subfigure}

%\vspace{-3mm}

  %\begin{subfigure}[t]{\textwidth}
    \centering
    \includegraphics[width=0.6\textwidth]{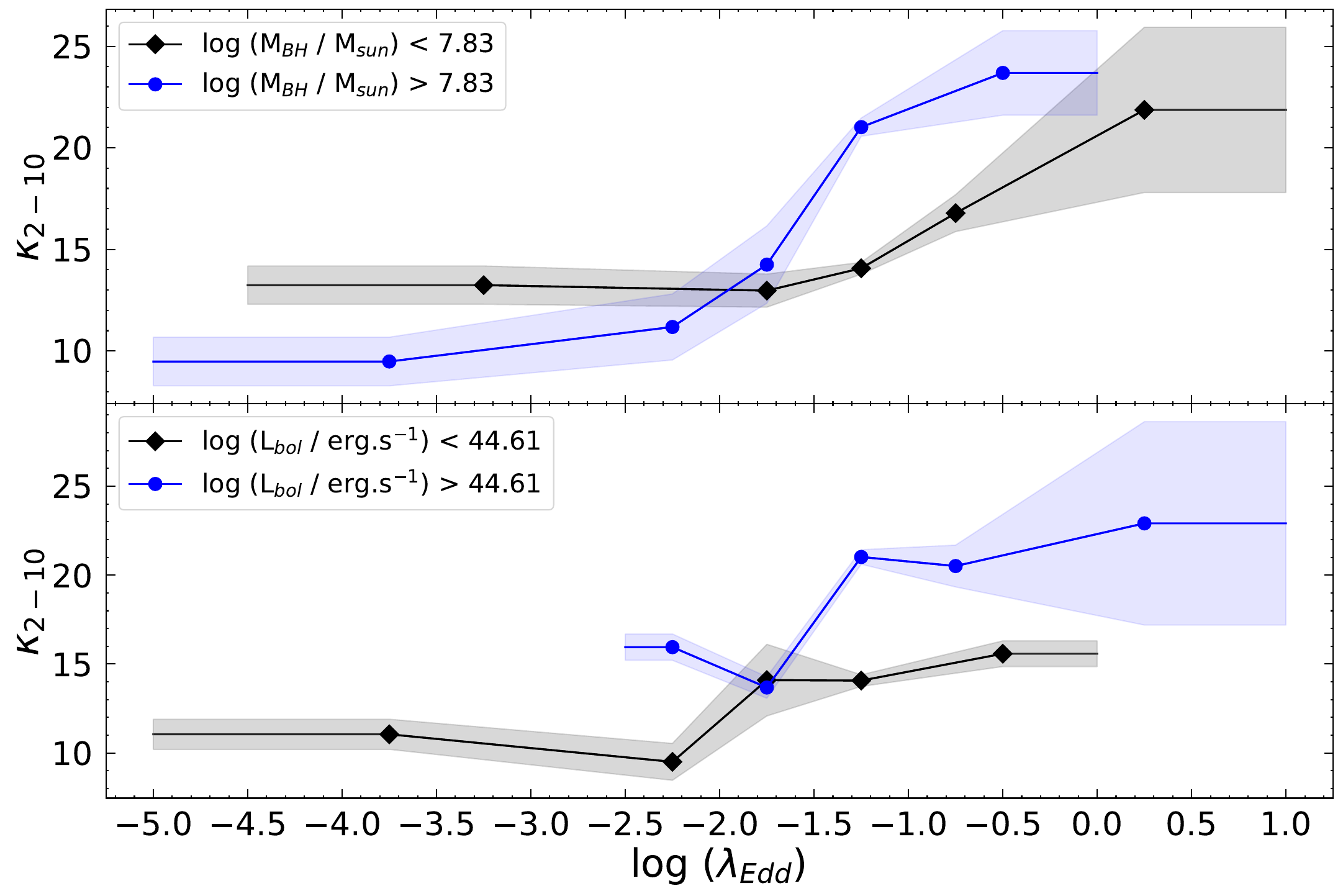}
    %\vspace{-65mm}
    %\caption{}
    \label{fig:kx_er_bin}
  %\end{subfigure}

%\vspace{-3mm}

  %\begin{subfigure}[t]{\textwidth}
    \centering
    \includegraphics[width=0.6\textwidth]{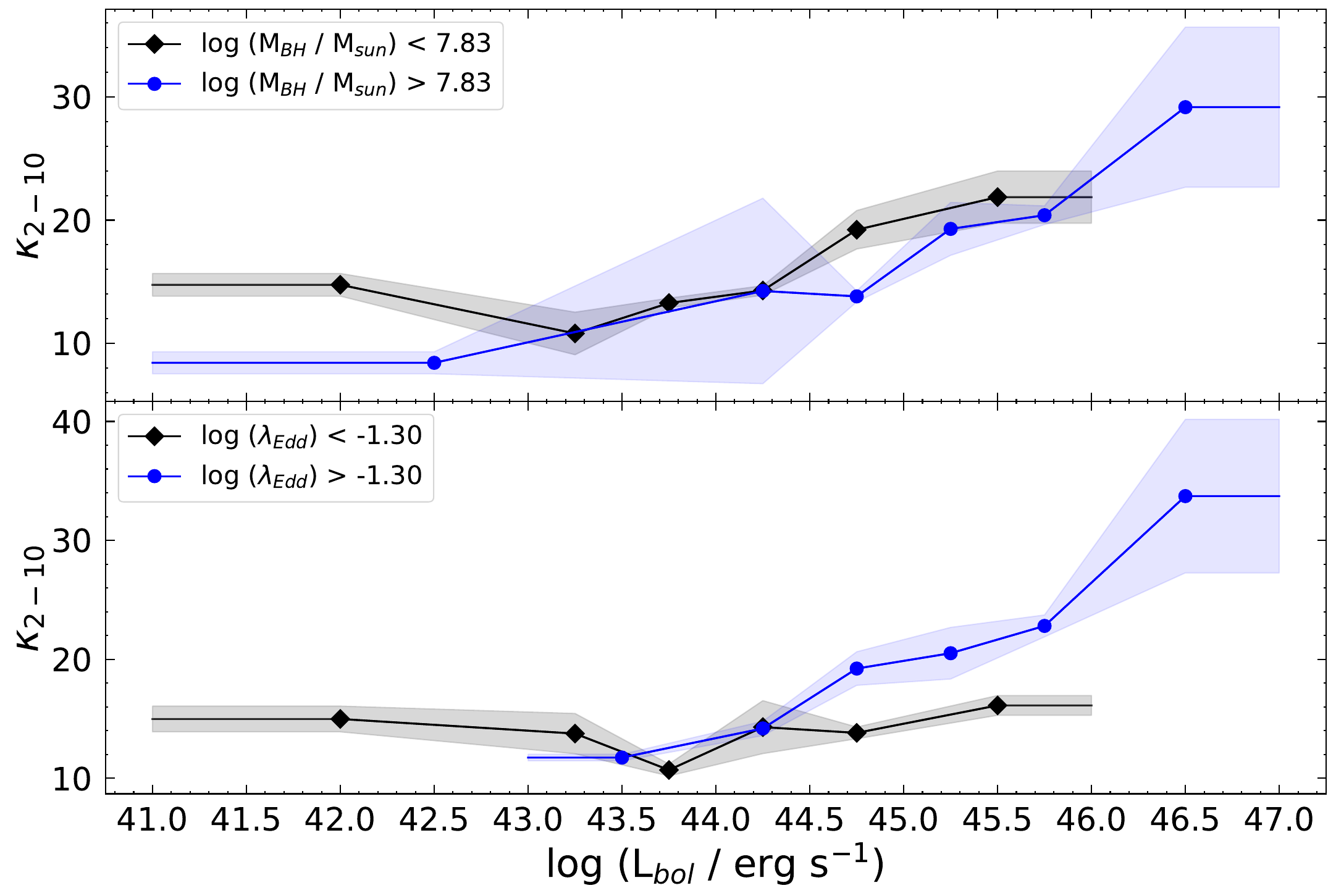}
    %\vspace{-65mm}
    %\caption{}
    \label{fig:kx_lb_bin}
  %\end{subfigure}
%\vspace{-3mm}
\caption{The 2--10\,keV bolometric correction vs AGN properties, for different ranges of bolometric luminosity, black hole mass, and Eddington ratio. We have plotted $\kappa_{2-10}$ as a function of $M_{\rm BH}$ (top panel) in bins of the bolometric luminosity (top) and Eddington ratio (bottom), $\lambda_{\rm Edd}$ (middle panel) in bins of the black hole mass (top) and the bolometric luminosity (bottom), and $L_{\rm bol}$ (bottom panel) in bins of black hole mass (top) and Eddington ratio (bottom). The black and blue lines show the median value of $\kappa_{2-10}$ in each bin of the corresponding AGN property, and the shaded gray and blue regions are the one sigma uncertainties on the median.}
\label{fig:kx_bin}
\end{figure*}

%------------------------------------------------------------------%
%------------------------------------------------------------------%

\subsection{What drives the X-ray Emission Fraction in AGN?}\label{sec:bc_bin}

Thanks to our large sample size spanning five orders of magnitude in luminosity, black hole mass, and Eddington ratio, we are able to divide our sample into bins of $M_{\rm BH}$, $L_{\rm bol}$, and $\lambda_{\rm Edd}$ to study their effects on the trends discussed above. In Fig. \ref{fig:kx_bin}, we plot $\kappa_{2-10}$ as a function of $M_{\rm BH}$, $\lambda_{\rm Edd}$, and $L_{\rm bol}$ in two bins (created at the median value) of the other two properties. The black diamonds and the blue circles correspond to the median value of $\kappa_{2-10}$ in the respective bin for each parameter's lower and higher values, respectively. The shaded regions denote the standard deviation around the median value in each bin. Even after dividing the sample into lower and higher value bins of bolometric luminosity (top panel in Fig. \ref{fig:kx_bin}), we do not find any significant dependence of $\kappa_{2-10}$ on $M_{\rm BH}$ ($p$-value = 0.64 and 0.23, respectively). However, when splitting the sample into higher and lower Eddington ratio bins, we uncover a weak but statistically-significant dependence ($p$-value = 0.007) of $\kappa_{2-10}$ on $M_{\rm BH}$ for ${\rm log}\,\lambda_{\rm Edd} > -1.3$. This dependence of $\kappa_{2-10}$ on $M_{\rm BH}$ only for higher Eddington ratios suggests a non-trivial role of the Eddington ratio in governing the X-ray bolometric corrections.

%------------------------------------------------------------------%
%------------------------------------------------------------------%

\begin{figure*}
\begin{center}
\includegraphics[width=1\textwidth]{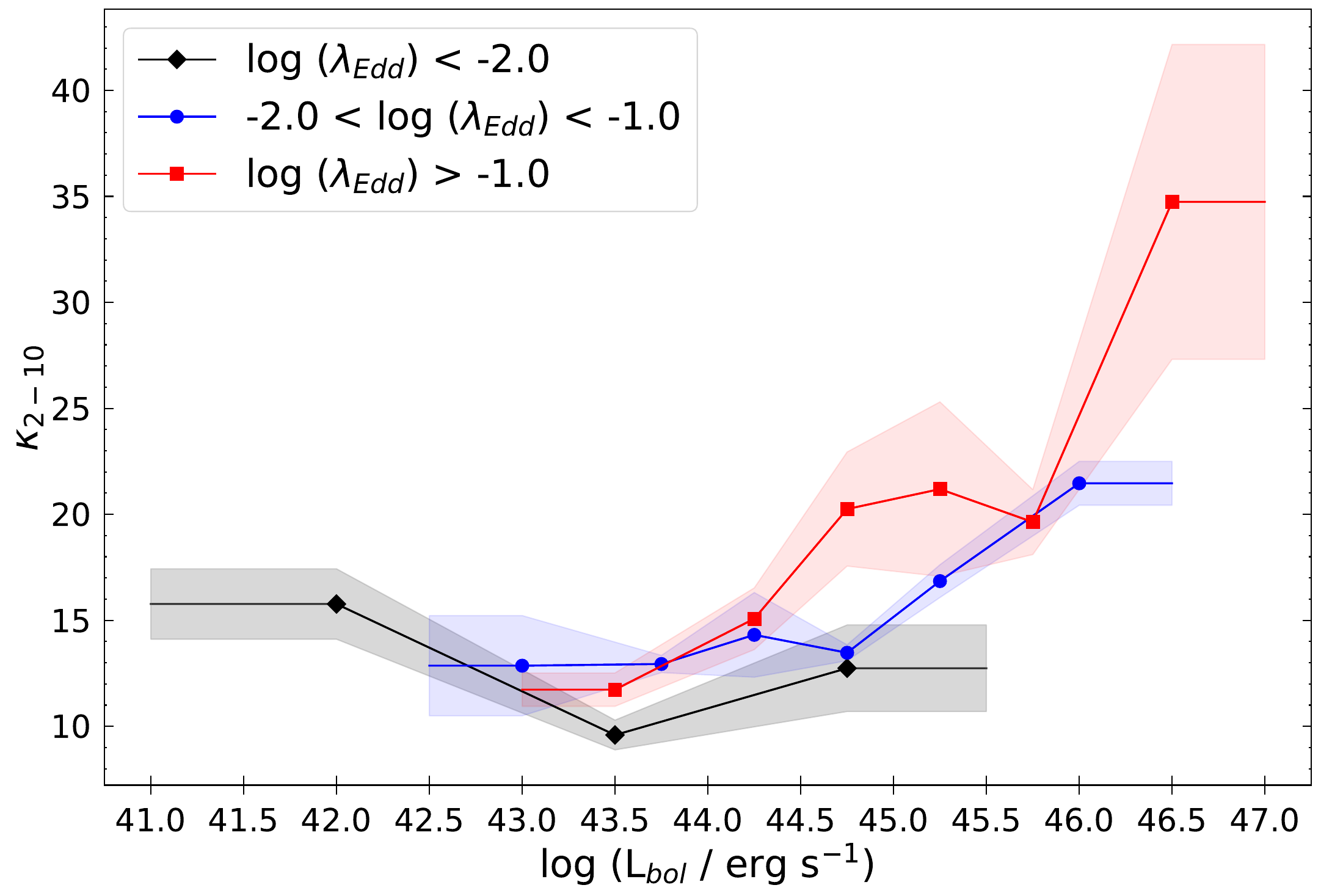}
 \caption{Relation between the 2--10\,keV bolometric correction and the bolometric luminosity in three bins of Eddington ratio. We show $\kappa_{2-10}$ as a function of $L_{\rm bol}$ for three different ranges of $\lambda_{\rm Edd}$. The black, blue, and red lines show the median value of $\kappa_{2-10}$ in each bin of $L_{\rm bol}$, for the three ranges of $\lambda_{\rm Edd}$ ($<0.01$, $0.01-0.1$, and $>0.1$). The shaded gray, blue, and red regions are the one sigma uncertainties on the median. The dependence of $\kappa_{2-10}$ on $L_{\rm bol}$ seems to disappear for $\lambda_{\rm Edd} < 0.01$.}
\label{fig:kx_lb_er_bin}
\end{center}
\end{figure*}
 
%------------------------------------------------------------------%
%------------------------------------------------------------------%

In the case of the Eddington ratio (Fig. \ref{fig:kx_bin}, middle panel), we recover the original increasing behavior of $\kappa_{2-10}$ for both higher and lower values of black hole mass and bolometric luminosity; further confirming the established dependence of $\kappa_{2-10}$ on $\lambda_{\rm Edd}$, irrespective of other parameters. Finally, we investigate the effects of black hole mass and Eddington ratio on the luminosity dependence of $\kappa_{2-10}$. Although the black hole mass does not affect the $L_{\rm bol}$--$\kappa_{2-10}$ correlation, we observe some dependence of this correlation on the Eddington ratio. The bottom panel of Fig. \ref{fig:kx_bin} indicates that for lower $\lambda_{\rm Edd}$ ($\lesssim 0.05$), the increase in $\kappa_{2-10}$ with $L_{\rm bol}$ completely disappears. This implies that the Eddington ratio somehow controls the dependence of $\kappa_{2-10}$ on the bolometric luminosity, and hence, would be a stronger regulator of the 2--10\,keV bolometric corrections. Therefore, the fraction of X-ray radiation in AGN is significantly dependent on the Eddington ratio, although luminosity may also have an important effect.

To further examine the effect of Eddington ratio on the $L_{\rm bol}$--$\kappa_{2-10}$ correlation, and to determine the exact value of $\lambda_{\rm Edd}$ where $\kappa_{2-10}$ shows a transition, we divide our sample into three ranges of Eddington ratio. As shown in Fig. \ref{fig:kx_lb_er_bin}, the black diamonds correspond to the $\lambda_{\rm Edd} < 0.01$ bin, the blue circles represent the sources with $0.01 < \lambda_{\rm Edd} < 0.1$, and the red squares are for the $\lambda_{\rm Edd} > 0.1$ bin. For each $L_{\rm bol}$ bin, we ensure that a minimum of six sources are included and then plot the median value of $\kappa_{2-10}$ in that bin. The resultant figure clearly shows an increment in $\kappa_{2-10}$ with $L_{\rm bol}$ for both the highest ($\lambda_{\rm Edd} > 0.1$) and intermediate ($0.01 < \lambda_{\rm Edd} < 0.1$) bins. We obtain an increase from almost $\kappa_{2-10}=10$ to 35 for the highest $\lambda_{\rm Edd}$ bin ($>0.1$). In the case of $\lambda_{\rm Edd}$ ranging between 0.01 and 0.1, $\kappa_{\rm 2-10}$ changes from around 12 up to 22 for a four orders of magnitude increase in bolometric luminosity. However, for the lowest Eddington ratio bin (below 0.01), the value of $\kappa_{2-10}$ oscillates between 10 and 15, not showing any significant positive trend with $L_{\rm bol}$. The lack of any correlation between $\kappa_{2-10}$ and $L_{\rm bol}$, specifically for $\lambda_{\rm Edd} \lesssim 0.01$, suggests a change in the AGN accretion flow around this value, such that the fraction of X-ray radiation emitted no longer correlates with the total luminosity of the AGN. Instead, this implies that, for low Eddington ratios, the X-ray emission scales with the total emission. This differs from what we observe in higher Eddington ratio and higher luminosity sources, where the relative contribution of the X-ray emission to the optical/UV decreases. This is also supported by a recent study of \citealp{2025A&A...697A..55K}, who used an X-ray illuminated disk model to show that the heating mechanism of the X-ray corona depends on the Eddington ratio, such that the fraction of power transferred to the corona decreases with increasing Eddington ratio.

%------------------------------------------------------------------%
%------------------------------------------------------------------%

\section{Discussion}\label{sec:discuss}

%------------------------------------------------------------------%
%------------------------------------------------------------------%

\subsection{Effect of Eddington Ratio on AGN Accretion}\label{sec:er}

The physical properties of the accretion flow in AGN are expected to change with the Eddington ratio (see \citealp{1999ASPC..161..295B} for a review). While the radiatively efficient, geometrically thin optically thick accretion disk model \citep{1973A&A....24..337S} is claimed to show good agreement with observations for AGN with intermediate Eddington ratios ($\lambda_{\rm Edd} \sim 0.01-1$; e.g., \citealp{2015MNRAS.446.3427C,2015MNRAS.449.4204L,2016MNRAS.460..212C}), it cannot reproduce the SEDs of systems with lower and higher Eddington ratios (e.g., \citealp{2012MNRAS.425..907J}). This is largely because these sources accrete at regimes that fall outside the validity range of the standard thin disk model and are likely dominated by radiatively inefficient accretion flows.

Several alternative accretion models have been proposed in the literature for such AGN, including radiatively inefficient slim disks for sources with high Eddington ratios (e.g., \citealp{2000PASJ...52..499M,2016MNRAS.456.3929S}) and advection-dominated accretion flows (ADAFs) for slowly accreting SMBHs (e.g., \citealp{1995ApJ...452..710N,2005Ap&SS.300..177N,2014ARA&A..52..529Y}). Specifically in the case of AGN with low accretion rates, our knowledge is very limited; mainly because these sources are relatively scarce, due to their low luminosities, even in our most complete and unbiased AGN surveys. Adding to their low detection rates, the study of these slow accreting LLAGN is even more challenging due to their emission properties, which are significantly different from the majority of the detected AGN population (concentrated around $0.01<\lambda_{\rm Edd}<0.1$). These sources usually lack a UV excess (e.g., \citealp{2008ARA&A..46..475H,2014MNRAS.438.2804N}) and almost always show a compact radio emission (e.g., \citealp{1998ApJ...497L..69H,2006ApJ...643..652N,2014MNRAS.438.2804N}).

%------------------------------------------------------------------%
%------------------------------------------------------------------%

\begin{figure*}
    %\begin{subfigure}[t]{0.49\textwidth}
    %\centering
    \includegraphics[width=0.49\textwidth]{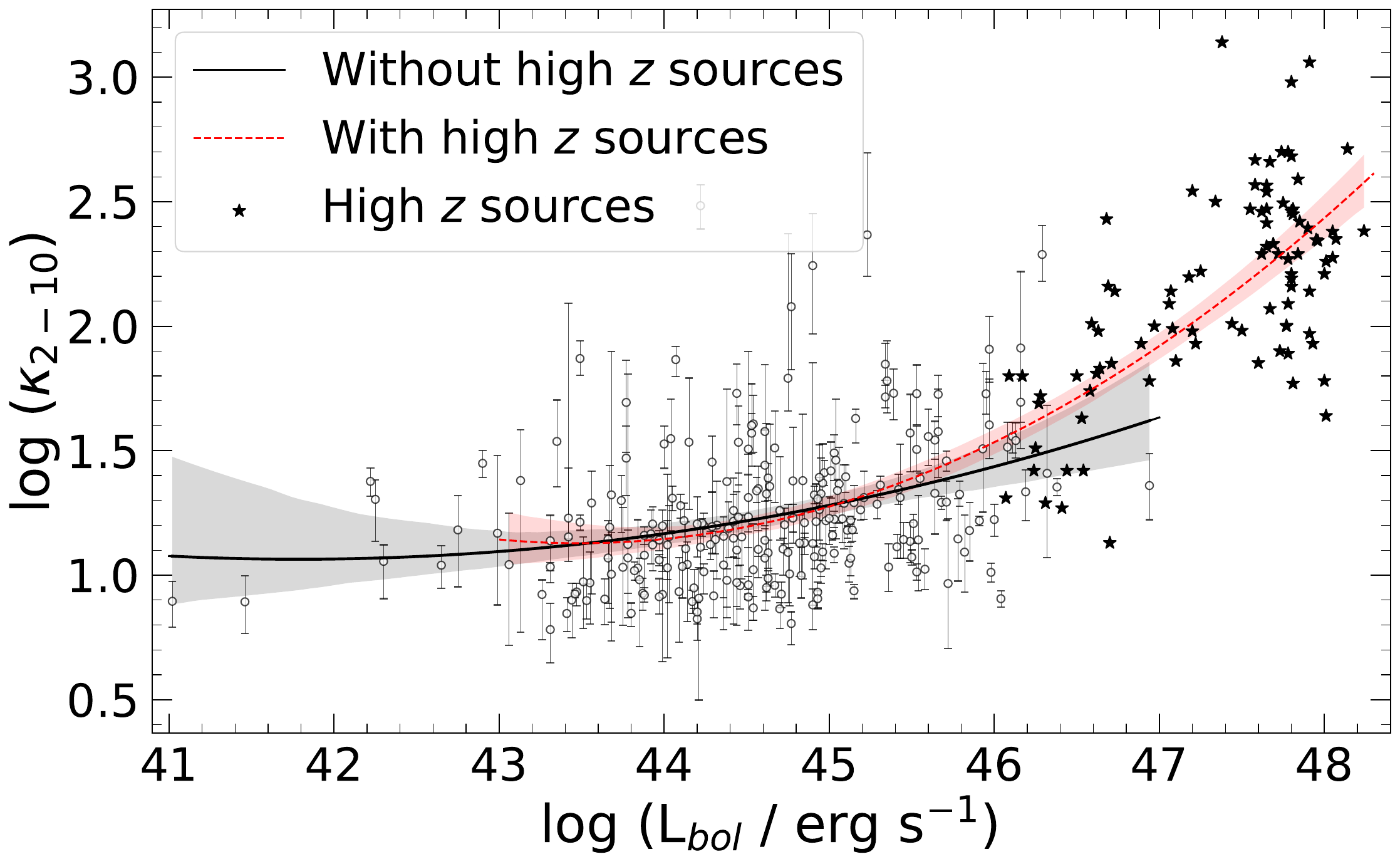}
    %\caption{}
    \label{fig:kx_lb_2}
    %\end{subfigure}
%\hspace{-1mm}
    %\begin{subfigure}[t]{0.49\textwidth}
    %\centering
    \includegraphics[width=0.49\textwidth]{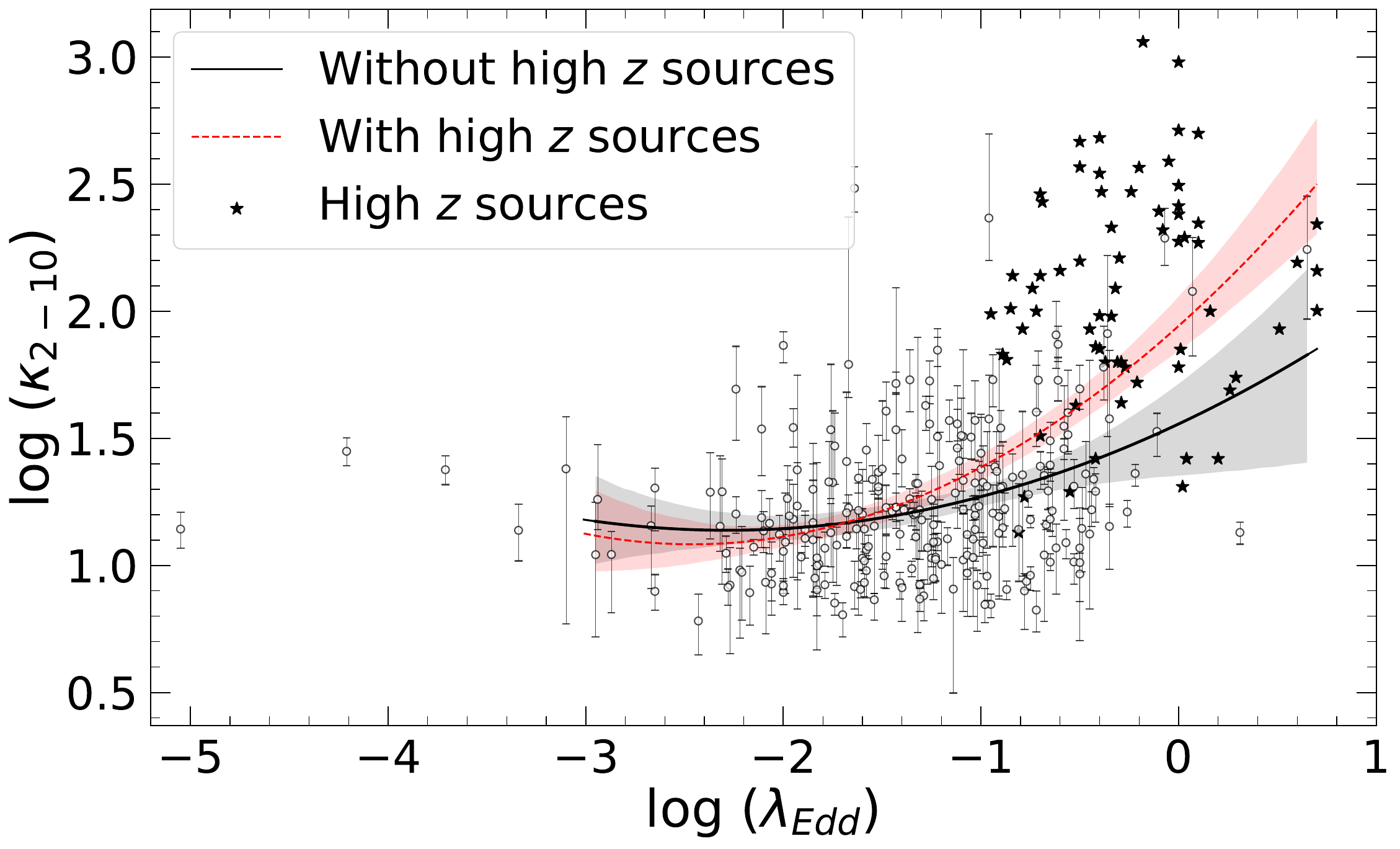}
    %\caption{}
    \label{fig:kx_er_2}
    %\end{subfigure}
    
\caption{Relation between the 2--10\,keV bolometric correction and AGN properties after including high-$z$ sources (shown as black stars) from \citet{2020A&A...636A..73D} and \citet{2023A&A...677A.111T}. Left panel: The best-fit relation (dashed red line) between $\kappa_{2-10}$ and $L_{\rm bol}$ after including high-$z$, high-$L_{\rm bol}$ sources shows a stronger increase at $L_{\rm bol} > 10^{46}\,\rm erg\,s^{-1}$ compared to our original fit (solid black line). Right panel: The dependence of $\kappa_{2-10}$ on $\lambda_{\rm Edd}$ (dashed red line), after including high-$z$, high-$\lambda_{\rm Edd}$ sources, is steeper than our original relation (solid black line).}
\label{fig:kx_2}
\end{figure*}

%------------------------------------------------------------------%
%------------------------------------------------------------------%

\subsection{Low Eddington Ratios}\label{sec:low_er}

Recent studies by \citet{2024MNRAS.534.2803H} and \citet{2025MNRAS.538..121K} of a sample of X-ray-selected unobscured AGN, using AGN SED models developed by \citet{2018MNRAS.480.1247K}, confirmed dramatic changes in the SED shapes of low accreting AGN ($\lambda_{\rm Edd} \lesssim 0.02$). They showed that the difference in the SED shape can be explained by the collapse of the accretion disk into an inefficient X-ray plasma, similar to what is observed in black hole X-ray binaries (BHXBs; \citealp{2007A&ARv..15....1D}). BHXBs generally show a change in their inner accretion flow at lower luminosities and accretion rates when the standard disk is replaced by an ADAF \citep{2003A&A...409..697M}. It seems that AGN SEDs might undergo a similar transition from a disk-dominated state to being dominated by the X-rays, around $\lambda_{\rm Edd} \sim 0.01-0.02$. This possible change with Eddington ratio in the accretion structure of AGN has been previously discussed in the context of \enquote*{changing-state} (CS) AGN, a class of AGN undergoing drastic changes in their spectral and emission properties (both in the optical/UV and X-rays) over short timescales (see \citealp{2023NatAs...7.1282R} for a review). \citet{2018MNRAS.480.3898N} showed that the decrease in the Eddington ratio, as the CS AGN entered a low accretion rate regime with $\lambda_{\rm Edd}<0.01$, was linked with a change in the spectral shape and possibly also in the accretion mechanism of these sources (e.g., \citealp{2025A&A...693A..35J}). Furthermore, the transition to low $\lambda_{\rm Edd}$ in these CS AGN is accompanied by a significant drop in the soft X-ray excess as the X-ray spectrum becomes harder. One would, in principle, expect this behavior to translate into the X-ray bolometric corrections for low Eddington ratio sources, which is exactly what we obtain from our analysis.

We clearly see a strong dependence of the established $\kappa_{2-10}-L_{\rm bol}$ relation on $\lambda_{\rm Edd}$ (Fig. \ref{fig:kx_lb_er_bin}). For sources with low Eddington ratios ($<0.01$), the flat trend between $\kappa_{2-10}$ and $L_{\rm bol}$ suggests that the X-ray emission increases with the bolometric luminosity. However, in the case of higher Eddington ratios (0.01--1), the X-ray emission saturates, while the optical/UV emission dominates the $L_{\rm bol}$. It is worth noting here that throughout our broadband SED modeling (described in detail in \citetalias{2024A&A...691A.203G}), we do not make any specific assumptions regarding the origin of the soft X-ray excess and only use the standard multi-temperature thin disk model to fit the optical/UV emission. However, our results are still consistent with other works that have used more complex models with a specific component to describe the soft X-ray excess (e.g., \citealp{2018MNRAS.480.3898N,2024MNRAS.534.2803H}). Therefore, our conclusions are not dependent on the model used to perform the SED fitting, further substantiating our claim that the Eddington ratio is indeed the main parameter driving the accretion and emission mechanisms in the nearby AGN population.

%------------------------------------------------------------------%
%------------------------------------------------------------------%

\subsection{High Eddington Ratios}\label{sec:high_er}

Based on our analysis, AGN with high Eddington ratios ($\lambda_{\rm Edd}>0.01$) show a clear increase in their X-ray bolometric corrections with luminosity, which could imply a saturation of their X-ray emission. However, since our sample is limited at very high luminosities ($L_{\rm bol} > 10^{46}\,\rm\,erg\,s^{-1}$) and Eddington ratios ($\lambda_{\rm Edd} > 1$), we cannot explore the behavior of $\kappa_{2-10}$ in these regimes. Therefore, we compiled AGN samples at higher redshifts to investigate the evolution of $\kappa_{2-10}$ at higher luminosities and Eddington ratios. To do so, we used the XXL ($0.9<z<5$) and WISSH ($2<z<4$) samples from \citet{2020A&A...636A..73D}, and the $z\sim3$ AGN sample from \citet{2023A&A...677A.111T}. The sources in these samples are extremely luminous with $L_{\rm bol} > 10^{46}\,\rm erg\,s^{-1}$ and a few of them have $\lambda_{\rm Edd} > 1$.

We show the dependence of $\kappa_{2-10}$ on $L_{\rm bol}$ and $\lambda_{\rm Edd}$ after including the high-$z$ sources in Fig.\,\ref{fig:kx_2}. For sources with $L_{\rm bol} = 10^{43}-10^{48}\,\rm erg\,s^{-1}$, the luminosity dependence of $\kappa_{2-10}$ can be described by the following equation:

\begin{equation}\label{eq:kx_lb_2}
    \begin{aligned}
        {\rm log}\,(\kappa_{2-10}) = &\,\,(122.18 \pm 19.13) + (-250.54 \pm 37.64)\\
        & \times \mathcal{L}_{\rm bol, 45} + (129.64 \pm 18.50) \times (\mathcal{L}_{\rm bol, 45})^2
    \end{aligned}
\end{equation}
Here, $\mathcal{L}_{\rm bol, 45} \equiv \log\,(L_{\rm bol} / 10^{45}\,\rm erg\,s^{-1})$. Inclusion of high-$L_{\rm bol}$ sources at higher redshifts shows that $\kappa_{2-10}$ increases significantly at $L_{\rm bol}>10^{46}\,\rm egs\,s^{-1}$ (Fig. \ref{fig:kx_2}, left panel). However, below that, this new relation (dashed red line) is consistent with our original best-fit relation for local AGN (solid black line). In the right panel of Fig. \ref{fig:kx_2}, we have plotted $\kappa_{2-10}$ as a function of $\lambda_{\rm Edd}$ after including the high-$z$ sources, and the second-order best-fit relation (for $\rm log\,\lambda_{\rm Edd} = -3$ to 1) is given as:

\begin{equation}\label{eq:kx_er_2}
    \begin{aligned}
        {\rm log}\,(\kappa_{2-10}) = &\,\,(1.94 \pm 0.04) + (0.70 \pm 0.08) \times {\rm log}\,(\lambda_{\rm Edd})\\
        & +(0.14 \pm 0.03) \times ({\rm log}\,[\lambda_{\rm Edd}])^2
    \end{aligned}
\end{equation}
Due to the addition of high-$z$ sources, the increase in $\kappa_{2-10}$ (dashed red line) with Eddington ratio is now steeper compared to what we obtained previously with only local sources (solid black line). However, it is important to note that this dependence could be due to sample selection effects at such high redshifts. We can clearly see an offset in the bolometric corrections from the high-$z$ sources compared to our local sample which is probably due to the high luminosities of the former. Therefore, it is necessary to clarify whether the high values of $\kappa_{2-10}$ are just due to the higher $L_{\rm bol}$ or it is actually an effect of the Eddington ratio. This can be achieved by having a similar redshift sample when comparing low and high Eddington ratio sources, which will be explored in a forthcoming publication (Kallova et al., in prep.)

%------------------------------------------------------------------%
%------------------------------------------------------------------%

\section{Summary and conclusions}\label{sec:summary}

This work focuses on determining the primary regulator of X-ray bolometric corrections using a large sample of hard-X-ray-selected nearby unobscured AGN, spanning five orders of magnitudes in bolometric luminosity, Eddington ratio, and black hole mass. We present results from the broadband SED modeling of 236 unobscured AGN with simultaneous optical, UV, and X-ray data (presented in \citetalias{2024A&A...691A.203G}). We estimated important quantities for our AGN sample, including bolometric luminosity, bolometric corrections, and Eddington ratio. We then investigated the dependence of the 2--10\,keV X-ray bolometric correction ($\kappa_{2-10}$) on the three main physical properties of AGN; $L_{\rm bol}$, $M_{\rm BH}$, and $\lambda_{\rm Edd}$. Here, we summarize our main findings and their implications on our understanding of AGN physics:

\begin{itemize}
    \item We obtain a positive correlation of $\kappa_{2-10}$ with the AGN bolometric luminosity (Eq. \ref{eq:kx_lb}, Fig. \ref{fig:kx}) and Eddington ratio (Eq. \ref{eq:kx_er}, Fig. \ref{fig:kx}), consistent with previous studies (e.g., \citealp{2020A&A...636A..73D}). This correlation persists for the entire range of black hole masses in our sample, extending from $\sim 10^{5} - 10^{10}\,M_{\odot}$.
    
    \item We do not find any significant dependence of $\kappa_{2-10}$ on the black hole mass for our AGN sample as a whole (Fig. \ref{fig:kx}). However, for higher Eddington ratio sources ($\lambda_{\rm Edd} > 0.05$), we obtain a weak correlation of $\kappa_{2-10}$ with $M_{\rm BH}$ (top panel of Fig. \ref{fig:kx_bin}).
    
    \item The Eddington ratio dependence of $\kappa_{2-10}$ is consistently recovered for all luminosity and black hole mass values in our sample (middle panel of Fig. \ref{fig:kx_bin}), with $L_{\rm bol}$ ranging from $10^{41}-10^{47}\,\rm erg\,s^{-1}$ and $M_{\rm BH} = 10^5 - 10^{10}\,M_{\odot}$. However, the well-established correlation between $\kappa_{2-10}$ and $L_{\rm bol}$ only exists for $\lambda_{\rm Edd} > 0.01$ and completely disappears for low Eddington ratios ($\lambda_{\rm Edd} < 0.01$; Fig. \ref{fig:kx_lb_er_bin}). This implies that between the bolometric luminosity and the Eddington ratio, the latter ($\lambda_{\rm Edd}$) is the primary regulator of the X-ray emission fraction in AGN. 

    \item Our analysis shows that the accretion mechanism of AGN depends on their Eddington ratio, such that the X-ray emission for low Eddington ratio AGN ($\lambda_{\rm Edd} < 0.01$) scales with the bolometric luminosity. However, for higher Eddington ratios ($\lambda_{\rm Edd} > 0.01$), the fraction of X-rays to optical/UV decreases.
    
    \item We also check how $\kappa_{2-10}$ behaves at higher luminosities ($L_{\rm bol} >10^{46}\,\rm erg\,s^{-1}$) and Eddington ratios ($\lambda_{\rm Edd} > 1$) by including high redshift sources in our study (Fig. \ref{fig:kx_2}). We find that $\kappa_{2-10}$ increases significantly with high luminosities and Eddington ratios (Eqs. \ref{eq:kx_lb_2} and \ref{eq:kx_er_2}). However, it should be noted that this could also be due to selection biases at high redshift.
    
 \end{itemize}

Over the years, many studies have reported the dependence of $\kappa_{2-10}$ on both $L_{\rm bol}$ and $\lambda_{\rm Edd}$. However, due to the complex interplay between the two quantities, still not completely understood, it was almost impossible to disentangle their effects on $\kappa_{2-10}$. Through the work presented here, we tried to examine separately the correlation of $\kappa_{2-10}$ with both of these parameters. We could conclusively show that the Eddington ratio drives the fraction of X-ray emission in AGN. Thanks to the hard X-ray selection criteria of the Swift/BAT AGN sample, it presents an unbiased and representative view of the AGN population in the local Universe. Hence, our results can be considered characteristic of the nearby unobscured AGN population. Moreover, since we use simultaneous optical, UV, and X-ray data to create the AGN SEDs and calculate their physical properties, we effectively reduced the effects of uncorrelated variability at different wavelengths on our analysis. However, this analysis can still be improved.

We acknowledge that much better and more sophisticated AGN models can be used to fit our broadband SEDs. Our main result clearly shows that low Eddington ratio AGN exhibits a different type of accretion flow than those with higher accretion rates. Using more complex and physically motivated models for such an analysis would be interesting. In an ongoing effort, we have used the \cite{2018MNRAS.480.1247K} AGN SED models to fit the broadband SEDs from our sample to interpret the effects of other parameters like black hole spin and the warm Comptonization component on the X-ray emission fraction in AGN (Kallov\'a et al., in prep.). We believe similar studies can help fill the gaps in our knowledge regarding the accretion physics of AGN over their entire demographics. Due to the lack of a proper theoretical framework to explain the observed trends between the multiwavelength emission of AGN, our understanding of the central engine of AGN is limited. Therefore, such observation-based studies from large, multiwavelength extragalactic surveys are pivotal in advancing AGN science.

%------------------------------------------------------------------%
%------------------------------------------------------------------%

\begin{acknowledgments}
    We thank the anonymous referee for their insightful comments that improved this manuscript. This work made use of data from the NASA/IPAC Infrared Science Archive and NASA/IPAC Extragalactic Database (NED), which are operated by the Jet Propulsion Laboratory, California Institute of Technology, under contract with the National Aeronautics and Space Administration. We acknowledge the usage of data and/or software provided by the High Energy Astrophysics Science Archive Research Center (HEASARC), which is a service of the Astrophysics Science Division at NASA/GSFC and the High Energy Astrophysics Division of the Smithsonian Astrophysical Observatory. We acknowledge financial support from: a 2018 grant from the ESO-Government of Chile Joint Committee (KKG); the Belgian Federal Science Policy Office (BELSPO) for the provision of financial support in the framework of the PRODEX Programme of the European Space Agency (KKG); ANID CATA-BASAL project FB210003 (CR, FEB, ET); ANID FONDECYT Regular grant \#1230345 (CR), \#1241005 (FEB, ET), \#1250821 (FEB, ET), and \#1200495 (ET); ANID Millennium Science Initiative - AIM23-0001 (FEB); NASA ADAP award 80NSSC19K0749 (MJK); Fondecyt Postdoctoral Fellowship \#3220516 (MJT), \#3230310 (YD), \#3230303 (AJ), and \#3230653 (IMC); STFC \#ST/X001075/1 (MJT). CR acknowledges support from the China-Chile joint research fund. BT acknowledges support from the European Research Council under the European Union's Horizon 2020 research and innovation program (grant agreement \#950533) and from the Israel Science Foundation (grant \#1849/19). This research was supported by the Excellence Cluster ORIGINS which is funded by the Deutsche Forschungsgemeinschaft (DFG, German Research Foundation) under Germany's Excellence Strategy - EXC 2094 - 390783311. BT also acknowledges the hospitality of the Instituto de Estudios Astrof\'isicos at Universidad Diego Portales, and of the Instituto de Astrof\'isica at Pontificia Universidad Cat\'olica de Chile. KO acknowledges support from the Korea Astronomy and Space Science Institute under the R\&D program (Project \#2025-1-831-01), supervised by the Korea AeroSpace Administration, and the National Research Foundation of Korea grant funded by the Korea government (MSIT; RS-2025-00553982).
\end{acknowledgments}

%------------------------------------------------------------------%
%------------------------------------------------------------------%

\bibliographystyle{aasjournal}
\bibliography{References} 

\begin{thebibliography}{}
\expandafter\ifx\csname natexlab\endcsname\relax\def\natexlab#1{#1}\fi
\providecommand{\url}[1]{\href{#1}{#1}}
\providecommand{\dodoi}[1]{doi:~\href{http://doi.org/#1}{\nolinkurl{#1}}}
\providecommand{\doeprint}[1]{\href{http://ascl.net/#1}{\nolinkurl{http://ascl.net/#1}}}
\providecommand{\doarXiv}[1]{\href{https://arxiv.org/abs/#1}{\nolinkurl{https://arxiv.org/abs/#1}}}

\bibitem[{{Avni} \& {Tananbaum}(1986)}]{1986ApJ...305...83A}
{Avni}, Y., \& {Tananbaum}, H. 1986, \apj, 305, 83, \dodoi{10.1086/164230}

\bibitem[{{Barthelmy} {et~al.}(2005){Barthelmy}, {Barbier}, {Cummings}, {Fenimore}, {Gehrels}, {Hullinger}, {Krimm}, {Markwardt}, {Palmer}, {Parsons}, {Sato}, {Suzuki}, {Takahashi}, {Tashiro}, \& {Tueller}}]{2005SSRv..120..143B}
{Barthelmy}, S.~D., {Barbier}, L.~M., {Cummings}, J.~R., {et~al.} 2005, \ssr, 120, 143, \dodoi{10.1007/s11214-005-5096-3}

\bibitem[{{Baumgartner} {et~al.}(2013){Baumgartner}, {Tueller}, {Markwardt}, {Skinner}, {Barthelmy}, {Mushotzky}, {Evans}, \& {Gehrels}}]{2013ApJS..207...19B}
{Baumgartner}, W.~H., {Tueller}, J., {Markwardt}, C.~B., {et~al.} 2013, \apjs, 207, 19, \dodoi{10.1088/0067-0049/207/2/19}

\bibitem[{{Bechtold} {et~al.}(1987){Bechtold}, {Czerny}, {Elvis}, {Fabbiano}, \& {Green}}]{1987ApJ...314..699B}
{Bechtold}, J., {Czerny}, B., {Elvis}, M., {Fabbiano}, G., \& {Green}, R.~F. 1987, \apj, 314, 699, \dodoi{10.1086/165098}

\bibitem[{{Beloborodov}(1999)}]{1999ASPC..161..295B}
{Beloborodov}, A.~M. 1999, in Astronomical Society of the Pacific Conference Series, Vol. 161, High Energy Processes in Accreting Black Holes, ed. J.~{Poutanen} \& R.~{Svensson}, 295, \dodoi{10.48550/arXiv.astro-ph/9901108}

\bibitem[{{Breeveld} {et~al.}(2010){Breeveld}, {Curran}, {Hoversten}, {Koch}, {Landsman}, {Marshall}, {Page}, {Poole}, {Roming}, {Smith}, {Still}, {Yershov}, {Blustin}, {Brown}, {Gronwall}, {Holland}, {Kuin}, {McGowan}, {Rosen}, {Boyd}, {Broos}, {Carter}, {Chester}, {Hancock}, {Huckle}, {Immler}, {Ivanushkina}, {Kennedy}, {Mason}, {Morgan}, {Oates}, {de Pasquale}, {Schady}, {Siegel}, \& {vand en Berk}}]{2010MNRAS.406.1687B}
{Breeveld}, A.~A., {Curran}, P.~A., {Hoversten}, E.~A., {et~al.} 2010, \mnras, 406, 1687, \dodoi{10.1111/j.1365-2966.2010.16832.x}

\bibitem[{{Burrows} {et~al.}(2005){Burrows}, {Hill}, {Nousek}, {Kennea}, {Wells}, {Osborne}, {Abbey}, {Beardmore}, {Mukerjee}, {Short}, {Chincarini}, {Campana}, {Citterio}, {Moretti}, {Pagani}, {Tagliaferri}, {Giommi}, {Capalbi}, {Tamburelli}, {Angelini}, {Cusumano}, {Br{\"a}uninger}, {Burkert}, \& {Hartner}}]{2005SSRv..120..165B}
{Burrows}, D.~N., {Hill}, J.~E., {Nousek}, J.~A., {et~al.} 2005, \ssr, 120, 165, \dodoi{10.1007/s11214-005-5097-2}

\bibitem[{{Capellupo} {et~al.}(2015){Capellupo}, {Netzer}, {Lira}, {Trakhtenbrot}, \& {Mej{\'\i}a-Restrepo}}]{2015MNRAS.446.3427C}
{Capellupo}, D.~M., {Netzer}, H., {Lira}, P., {Trakhtenbrot}, B., \& {Mej{\'\i}a-Restrepo}, J. 2015, \mnras, 446, 3427, \dodoi{10.1093/mnras/stu2266}

\bibitem[{{Capellupo} {et~al.}(2016){Capellupo}, {Netzer}, {Lira}, {Trakhtenbrot}, \& {Mej{\'\i}a-Restrepo}}]{2016MNRAS.460..212C}
---. 2016, \mnras, 460, 212, \dodoi{10.1093/mnras/stw937}

\bibitem[{{Collier} \& {Peterson}(2001)}]{2001ApJ...555..775C}
{Collier}, S., \& {Peterson}, B.~M. 2001, \apj, 555, 775, \dodoi{10.1086/321517}

\bibitem[{{Crummy} {et~al.}(2006){Crummy}, {Fabian}, {Gallo}, \& {Ross}}]{2006MNRAS.365.1067C}
{Crummy}, J., {Fabian}, A.~C., {Gallo}, L., \& {Ross}, R.~R. 2006, \mnras, 365, 1067, \dodoi{10.1111/j.1365-2966.2005.09844.x}

\bibitem[{{Done} {et~al.}(2012){Done}, {Davis}, {Jin}, {Blaes}, \& {Ward}}]{2012MNRAS.420.1848D}
{Done}, C., {Davis}, S.~W., {Jin}, C., {Blaes}, O., \& {Ward}, M. 2012, \mnras, 420, 1848, \dodoi{10.1111/j.1365-2966.2011.19779.x}

\bibitem[{{Done} {et~al.}(2007){Done}, {Gierli{\'n}ski}, \& {Kubota}}]{2007A&ARv..15....1D}
{Done}, C., {Gierli{\'n}ski}, M., \& {Kubota}, A. 2007, \aapr, 15, 1, \dodoi{10.1007/s00159-007-0006-1}

\bibitem[{{Duras} {et~al.}(2020){Duras}, {Bongiorno}, {Ricci}, {Piconcelli}, {Shankar}, {Lusso}, {Bianchi}, {Fiore}, {Maiolino}, {Marconi}, {Onori}, {Sani}, {Schneider}, {Vignali}, \& {La Franca}}]{2020A&A...636A..73D}
{Duras}, F., {Bongiorno}, A., {Ricci}, F., {et~al.} 2020, \aap, 636, A73, \dodoi{10.1051/0004-6361/201936817}

\bibitem[{{Garc{\'\i}a} {et~al.}(2019){Garc{\'\i}a}, {Kara}, {Walton}, {Beuchert}, {Dauser}, {Gatuzz}, {Balokovic}, {Steiner}, {Tombesi}, {Connors}, {Kallman}, {Harrison}, {Fabian}, {Wilms}, {Stern}, {Lanz}, {Ricci}, \& {Ballantyne}}]{2019ApJ...871...88G}
{Garc{\'\i}a}, J.~A., {Kara}, E., {Walton}, D., {et~al.} 2019, \apj, 871, 88, \dodoi{10.3847/1538-4357/aaf739}

\bibitem[{{Graham} {et~al.}(2020){Graham}, {Ross}, {Stern}, {Drake}, {McKernan}, {Ford}, {Djorgovski}, {Mahabal}, {Glikman}, {Larson}, \& {Christensen}}]{2020MNRAS.491.4925G}
{Graham}, M.~J., {Ross}, N.~P., {Stern}, D., {et~al.} 2020, \mnras, 491, 4925, \dodoi{10.1093/mnras/stz3244}

\bibitem[{{Gupta} {et~al.}(2024){Gupta}, {Ricci}, {Temple}, {Tortosa}, {Koss}, {Assef}, {Bauer}, {Mushotzy}, {Ricci}, {Ueda}, {Rojas}, {Trakhtenbrot}, {Chang}, {Oh}, {Li}, {Kawamuro}, {Diaz}, {Powell}, {Stern}, {Megan Urry}, {Harrison}, \& {Cenko}}]{2024A&A...691A.203G}
{Gupta}, K.~K., {Ricci}, C., {Temple}, M.~J., {et~al.} 2024, \aap, 691, A203, \dodoi{10.1051/0004-6361/202450567}

\bibitem[{{Haardt} \& {Maraschi}(1991)}]{1991ApJ...380L..51H}
{Haardt}, F., \& {Maraschi}, L. 1991, \apjl, 380, L51, \dodoi{10.1086/186171}

\bibitem[{{Hagen} {et~al.}(2024){Hagen}, {Done}, {Silverman}, {Li}, {Liu}, {Ren}, {Buchner}, {Merloni}, {Nagao}, \& {Salvato}}]{2024MNRAS.534.2803H}
{Hagen}, S., {Done}, C., {Silverman}, J.~D., {et~al.} 2024, \mnras, 534, 2803, \dodoi{10.1093/mnras/stae2272}

\bibitem[{{Herrnstein} {et~al.}(1998){Herrnstein}, {Greenhill}, {Moran}, {Diamond}, {Inoue}, {Nakai}, \& {Miyoshi}}]{1998ApJ...497L..69H}
{Herrnstein}, J.~R., {Greenhill}, L.~J., {Moran}, J.~M., {et~al.} 1998, \apjl, 497, L69, \dodoi{10.1086/311284}

\bibitem[{{Ho}(2008)}]{2008ARA&A..46..475H}
{Ho}, L.~C. 2008, \araa, 46, 475, \dodoi{10.1146/annurev.astro.45.051806.110546}

\bibitem[{{Hopkins} {et~al.}(2024){Hopkins}, {Squire}, {Quataert}, {Murray}, {Su}, {Steinwandel}, {Kremer}, {Faucher-Giguere}, \& {Wellons}}]{2024OJAp....7E..20H}
{Hopkins}, P.~F., {Squire}, J., {Quataert}, E., {et~al.} 2024, The Open Journal of Astrophysics, 7, 20, \dodoi{10.21105/astro.2310.04507}

\bibitem[{{Jana} {et~al.}(2025){Jana}, {Ricci}, {Temple}, {Chang}, {Shablovinskaya}, {Trakhtenbrot}, {Diaz}, {Ilic}, {Nandi}, \& {Koss}}]{2025A&A...693A..35J}
{Jana}, A., {Ricci}, C., {Temple}, M.~J., {et~al.} 2025, \aap, 693, A35, \dodoi{10.1051/0004-6361/202451058}

\bibitem[{{Jin} {et~al.}(2012){Jin}, {Ward}, \& {Done}}]{2012MNRAS.425..907J}
{Jin}, C., {Ward}, M., \& {Done}, C. 2012, \mnras, 425, 907, \dodoi{10.1111/j.1365-2966.2012.21272.x}

\bibitem[{{Kammoun} {et~al.}(2025){Kammoun}, {Papadakis}, {Dov{\v{c}}iak}, {Lusso}, {Nardini}, \& {Risaliti}}]{2025A&A...697A..55K}
{Kammoun}, E., {Papadakis}, I.~E., {Dov{\v{c}}iak}, M., {et~al.} 2025, \aap, 697, A55, \dodoi{10.1051/0004-6361/202452629}

\bibitem[{{Kang} {et~al.}(2025){Kang}, {Done}, {Hagen}, {Temple}, {Silverman}, {Li}, \& {Liu}}]{2025MNRAS.538..121K}
{Kang}, J.-L., {Done}, C., {Hagen}, S., {et~al.} 2025, \mnras, 538, 121, \dodoi{10.1093/mnras/staf145}

\bibitem[{{Kelly}(2007)}]{2007ApJ...665.1489K}
{Kelly}, B.~C. 2007, \apj, 665, 1489, \dodoi{10.1086/519947}

\bibitem[{{Koratkar} \& {Blaes}(1999)}]{1999PASP..111....1K}
{Koratkar}, A., \& {Blaes}, O. 1999, \pasp, 111, 1, \dodoi{10.1086/316294}

\bibitem[{{Koss} {et~al.}(2017){Koss}, {Trakhtenbrot}, {Ricci}, {Lamperti}, {Oh}, {Berney}, {Schawinski}, {Balokovi{\'c}}, {Baronchelli}, {Crenshaw}, {Fischer}, {Gehrels}, {Harrison}, {Hashimoto}, {Hogg}, {Ichikawa}, {Masetti}, {Mushotzky}, {Sartori}, {Stern}, {Treister}, {Ueda}, {Veilleux}, \& {Winter}}]{2017ApJ...850...74K}
{Koss}, M., {Trakhtenbrot}, B., {Ricci}, C., {et~al.} 2017, \apj, 850, 74, \dodoi{10.3847/1538-4357/aa8ec9}

\bibitem[{{Koss} {et~al.}(2022){Koss}, {Ricci}, {Trakhtenbrot}, {Oh}, {den Brok}, {Mej{\'\i}a-Restrepo}, {Stern}, {Privon}, {Treister}, {Powell}, {Mushotzky}, {Bauer}, {Ananna}, {Balokovi{\'c}}, {B{\"a}r}, {Becker}, {Bessiere}, {Burtscher}, {Caglar}, {Congiu}, {Evans}, {Harrison}, {Heida}, {Ichikawa}, {Kamraj}, {Lamperti}, {Pacucci}, {Ricci}, {Riffel}, {Rojas}, {Schawinski}, {Temple}, {Urry}, {Veilleux}, \& {Williams}}]{2022ApJS..261....2K}
{Koss}, M.~J., {Ricci}, C., {Trakhtenbrot}, B., {et~al.} 2022, \apjs, 261, 2, \dodoi{10.3847/1538-4365/ac6c05}

\bibitem[{{Krimm} {et~al.}(2013){Krimm}, {Holland}, {Corbet}, {Pearlman}, {Romano}, {Kennea}, {Bloom}, {Barthelmy}, {Baumgartner}, {Cummings}, {Gehrels}, {Lien}, {Markwardt}, {Palmer}, {Sakamoto}, {Stamatikos}, \& {Ukwatta}}]{2013ApJS..209...14K}
{Krimm}, H.~A., {Holland}, S.~T., {Corbet}, R.~H.~D., {et~al.} 2013, \apjs, 209, 14, \dodoi{10.1088/0067-0049/209/1/14}

\bibitem[{{Kubota} \& {Done}(2018)}]{2018MNRAS.480.1247K}
{Kubota}, A., \& {Done}, C. 2018, \mnras, 480, 1247, \dodoi{10.1093/mnras/sty1890}

\bibitem[{{LaMassa} {et~al.}(2015){LaMassa}, {Cales}, {Moran}, {Myers}, {Richards}, {Eracleous}, {Heckman}, {Gallo}, \& {Urry}}]{2015ApJ...800..144L}
{LaMassa}, S.~M., {Cales}, S., {Moran}, E.~C., {et~al.} 2015, \apj, 800, 144, \dodoi{10.1088/0004-637X/800/2/144}

\bibitem[{{Lawrence}(2018)}]{2018NatAs...2..102L}
{Lawrence}, A. 2018, Nature Astronomy, 2, 102, \dodoi{10.1038/s41550-017-0372-1}

\bibitem[{{Lusso} \& {Risaliti}(2016)}]{2016ApJ...819..154L}
{Lusso}, E., \& {Risaliti}, G. 2016, \apj, 819, 154, \dodoi{10.3847/0004-637X/819/2/154}

\bibitem[{{Lusso} \& {Risaliti}(2017)}]{2017A&A...602A..79L}
---. 2017, \aap, 602, A79, \dodoi{10.1051/0004-6361/201630079}

\bibitem[{{Lusso} {et~al.}(2015){Lusso}, {Worseck}, {Hennawi}, {Prochaska}, {Vignali}, {Stern}, \& {O'Meara}}]{2015MNRAS.449.4204L}
{Lusso}, E., {Worseck}, G., {Hennawi}, J.~F., {et~al.} 2015, \mnras, 449, 4204, \dodoi{10.1093/mnras/stv516}

\bibitem[{{Lusso} {et~al.}(2012){Lusso}, {Comastri}, {Simmons}, {Mignoli}, {Zamorani}, {Vignali}, {Brusa}, {Shankar}, {Lutz}, {Trump}, {Maiolino}, {Gilli}, {Bolzonella}, {Puccetti}, {Salvato}, {Impey}, {Civano}, {Elvis}, {Mainieri}, {Silverman}, {Koekemoer}, {Bongiorno}, {Merloni}, {Berta}, {Le Floc'h}, {Magnelli}, {Pozzi}, \& {Riguccini}}]{2012MNRAS.425..623L}
{Lusso}, E., {Comastri}, A., {Simmons}, B.~D., {et~al.} 2012, \mnras, 425, 623, \dodoi{10.1111/j.1365-2966.2012.21513.x}

\bibitem[{{Lynden-Bell}(1969)}]{1969Natur.223..690L}
{Lynden-Bell}, D. 1969, \nat, 223, 690, \dodoi{10.1038/223690a0}

\bibitem[{{Maccarone}(2003)}]{2003A&A...409..697M}
{Maccarone}, T.~J. 2003, \aap, 409, 697, \dodoi{10.1051/0004-6361:20031146}

\bibitem[{{Magdziarz} {et~al.}(1998){Magdziarz}, {Blaes}, {Zdziarski}, {Johnson}, \& {Smith}}]{1998MNRAS.301..179M}
{Magdziarz}, P., {Blaes}, O.~M., {Zdziarski}, A.~A., {Johnson}, W.~N., \& {Smith}, D.~A. 1998, \mnras, 301, 179, \dodoi{10.1046/j.1365-8711.1998.02015.x}

\bibitem[{{Marconi} {et~al.}(2004){Marconi}, {Risaliti}, {Gilli}, {Hunt}, {Maiolino}, \& {Salvati}}]{2004IAUS..222...49M}
{Marconi}, A., {Risaliti}, G., {Gilli}, R., {et~al.} 2004, in IAU Symposium, Vol. 222, The Interplay Among Black Holes, Stars and ISM in Galactic Nuclei, ed. T.~{Storchi-Bergmann}, L.~C. {Ho}, \& H.~R. {Schmitt}, 49--52, \dodoi{10.1017/S1743921304001437}

\bibitem[{{Mej{\'\i}a-Restrepo} {et~al.}(2022){Mej{\'\i}a-Restrepo}, {Trakhtenbrot}, {Koss}, {Oh}, {den Brok}, {Stern}, {Powell}, {Ricci}, {Caglar}, {Ricci}, {Bauer}, {Treister}, {Harrison}, {Urry}, {Ananna}, {Asmus}, {Assef}, {B{\"a}r}, {Bessiere}, {Burtscher}, {Ichikawa}, {Kakkad}, {Kamraj}, {Mushotzky}, {Privon}, {Rojas}, {Sani}, {Schawinski}, \& {Veilleux}}]{2022ApJS..261....5M}
{Mej{\'\i}a-Restrepo}, J.~E., {Trakhtenbrot}, B., {Koss}, M.~J., {et~al.} 2022, \apjs, 261, 5, \dodoi{10.3847/1538-4365/ac6602}

\bibitem[{{Mineshige} {et~al.}(2000){Mineshige}, {Kawaguchi}, {Takeuchi}, \& {Hayashida}}]{2000PASJ...52..499M}
{Mineshige}, S., {Kawaguchi}, T., {Takeuchi}, M., \& {Hayashida}, K. 2000, \pasj, 52, 499, \dodoi{10.1093/pasj/52.3.499}

\bibitem[{{Narayan}(2005)}]{2005Ap&SS.300..177N}
{Narayan}, R. 2005, \apss, 300, 177, \dodoi{10.1007/s10509-005-1178-7}

\bibitem[{{Narayan} \& {Yi}(1995)}]{1995ApJ...452..710N}
{Narayan}, R., \& {Yi}, I. 1995, \apj, 452, 710, \dodoi{10.1086/176343}

\bibitem[{{Nemmen} {et~al.}(2014){Nemmen}, {Storchi-Bergmann}, \& {Eracleous}}]{2014MNRAS.438.2804N}
{Nemmen}, R.~S., {Storchi-Bergmann}, T., \& {Eracleous}, M. 2014, \mnras, 438, 2804, \dodoi{10.1093/mnras/stt2388}

\bibitem[{{Nemmen} {et~al.}(2006){Nemmen}, {Storchi-Bergmann}, {Yuan}, {Eracleous}, {Terashima}, \& {Wilson}}]{2006ApJ...643..652N}
{Nemmen}, R.~S., {Storchi-Bergmann}, T., {Yuan}, F., {et~al.} 2006, \apj, 643, 652, \dodoi{10.1086/500571}

\bibitem[{{Netzer}(2019)}]{2019MNRAS.488.5185N}
{Netzer}, H. 2019, \mnras, 488, 5185, \dodoi{10.1093/mnras/stz2016}

\bibitem[{{Noda} \& {Done}(2018)}]{2018MNRAS.480.3898N}
{Noda}, H., \& {Done}, C. 2018, \mnras, 480, 3898, \dodoi{10.1093/mnras/sty2032}

\bibitem[{{Novikov} \& {Thorne}(1973)}]{1973blho.conf..343N}
{Novikov}, I.~D., \& {Thorne}, K.~S. 1973, in Black Holes (Les Astres Occlus), 343--450

\bibitem[{{Panagiotou} {et~al.}(2022){Panagiotou}, {Papadakis}, {Kara}, {Kammoun}, \& {Dov{\v{c}}iak}}]{2022ApJ...935...93P}
{Panagiotou}, C., {Papadakis}, I., {Kara}, E., {Kammoun}, E., \& {Dov{\v{c}}iak}, M. 2022, \apj, 935, 93, \dodoi{10.3847/1538-4357/ac7e4d}

\bibitem[{{Peng} {et~al.}(2002){Peng}, {Ho}, {Impey}, \& {Rix}}]{2002AJ....124..266P}
{Peng}, C.~Y., {Ho}, L.~C., {Impey}, C.~D., \& {Rix}, H.-W. 2002, \aj, 124, 266, \dodoi{10.1086/340952}

\bibitem[{{Peng} {et~al.}(2010){Peng}, {Ho}, {Impey}, \& {Rix}}]{2010AJ....139.2097P}
---. 2010, \aj, 139, 2097, \dodoi{10.1088/0004-6256/139/6/2097}

\bibitem[{{Petrucci} {et~al.}(2018){Petrucci}, {Ursini}, {De Rosa}, {Bianchi}, {Cappi}, {Matt}, {Dadina}, \& {Malzac}}]{2018A&A...611A..59P}
{Petrucci}, P.~O., {Ursini}, F., {De Rosa}, A., {et~al.} 2018, \aap, 611, A59, \dodoi{10.1051/0004-6361/201731580}

\bibitem[{{Poole} {et~al.}(2008){Poole}, {Breeveld}, {Page}, {Land sman}, {Holland}, {Roming}, {Kuin}, {Brown}, {Gronwall}, {Hunsberger}, {Koch}, {Mason}, {Schady}, {vanden Berk}, {Blustin}, {Boyd}, {Broos}, {Carter}, {Chester}, {Cucchiara}, {Hancock}, {Huckle}, {Immler}, {Ivanushkina}, {Kennedy}, {Marshall}, {Morgan}, {Pandey}, {de Pasquale}, {Smith}, \& {Still}}]{2008MNRAS.383..627P}
{Poole}, T.~S., {Breeveld}, A.~A., {Page}, M.~J., {et~al.} 2008, \mnras, 383, 627, \dodoi{10.1111/j.1365-2966.2007.12563.x}

\bibitem[{{Rees}(1984)}]{1984ARA&A..22..471R}
{Rees}, M.~J. 1984, \araa, 22, 471, \dodoi{10.1146/annurev.aa.22.090184.002351}

\bibitem[{{Ricci} \& {Trakhtenbrot}(2023)}]{2023NatAs...7.1282R}
{Ricci}, C., \& {Trakhtenbrot}, B. 2023, Nature Astronomy, 7, 1282, \dodoi{10.1038/s41550-023-02108-4}

\bibitem[{{Ricci} {et~al.}(2015){Ricci}, {Ueda}, {Koss}, {Trakhtenbrot}, {Bauer}, \& {Gandhi}}]{2015ApJ...815L..13R}
{Ricci}, C., {Ueda}, Y., {Koss}, M.~J., {et~al.} 2015, \apjl, 815, L13, \dodoi{10.1088/2041-8205/815/1/L13}

\bibitem[{{Ricci} {et~al.}(2017){Ricci}, {Trakhtenbrot}, {Koss}, {Ueda}, {Del Vecchio}, {Treister}, {Schawinski}, {Paltani}, {Oh}, {Lamperti}, {Berney}, {Gand hi}, {Ichikawa}, {Bauer}, {Ho}, {Asmus}, {Beckmann}, {Soldi}, {Balokovi{\'c}}, {Gehrels}, \& {Markwardt}}]{2017ApJS..233...17R}
{Ricci}, C., {Trakhtenbrot}, B., {Koss}, M.~J., {et~al.} 2017, \apjs, 233, 17, \dodoi{10.3847/1538-4365/aa96ad}

\bibitem[{{Rumbaugh} {et~al.}(2018){Rumbaugh}, {Shen}, {Morganson}, {Liu}, {Banerji}, {McMahon}, {Abdalla}, {Benoit-L{\'e}vy}, {Bertin}, {Brooks}, {Buckley-Geer}, {Capozzi}, {Carnero Rosell}, {Carrasco Kind}, {Carretero}, {Cunha}, {D'Andrea}, {da Costa}, {DePoy}, {Desai}, {Doel}, {Frieman}, {Garc{\'\i}a-Bellido}, {Gruen}, {Gruendl}, {Gschwend}, {Gutierrez}, {Honscheid}, {James}, {Kuehn}, {Kuhlmann}, {Kuropatkin}, {Lima}, {Maia}, {Marshall}, {Martini}, {Menanteau}, {Plazas}, {Reil}, {Roodman}, {Sanchez}, {Scarpine}, {Schindler}, {Schubnell}, {Sheldon}, {Smith}, {Soares-Santos}, {Sobreira}, {Suchyta}, {Swanson}, {Walker}, {Wester}, \& {DES Collaboration}}]{2018ApJ...854..160R}
{Rumbaugh}, N., {Shen}, Y., {Morganson}, E., {et~al.} 2018, \apj, 854, 160, \dodoi{10.3847/1538-4357/aaa9b6}

\bibitem[{{Salpeter}(1964)}]{1964ApJ...140..796S}
{Salpeter}, E.~E. 1964, \apj, 140, 796, \dodoi{10.1086/147973}

\bibitem[{{Shakura} \& {Sunyaev}(1973)}]{1973A&A....24..337S}
{Shakura}, N.~I., \& {Sunyaev}, R.~A. 1973, \aap, 500, 33

\bibitem[{{S{\k{a}}dowski} \& {Narayan}(2016)}]{2016MNRAS.456.3929S}
{S{\k{a}}dowski}, A., \& {Narayan}, R. 2016, \mnras, 456, 3929, \dodoi{10.1093/mnras/stv2941}

\bibitem[{{Steffen} {et~al.}(2006){Steffen}, {Strateva}, {Brandt}, {Alexand er}, {Koekemoer}, {Lehmer}, {Schneider}, \& {Vignali}}]{2006AJ....131.2826S}
{Steffen}, A.~T., {Strateva}, I., {Brandt}, W.~N., {et~al.} 2006, \aj, 131, 2826, \dodoi{10.1086/503627}

\bibitem[{{Trefoloni} {et~al.}(2023){Trefoloni}, {Lusso}, {Nardini}, {Risaliti}, {Bargiacchi}, {Bisogni}, {Civano}, {Elvis}, {Fabbiano}, {Gilli}, {Marconi}, {Richards}, {Sacchi}, {Salvestrini}, {Signorini}, \& {Vignali}}]{2023A&A...677A.111T}
{Trefoloni}, B., {Lusso}, E., {Nardini}, E., {et~al.} 2023, \aap, 677, A111, \dodoi{10.1051/0004-6361/202346024}

\bibitem[{{Vasudevan} \& {Fabian}(2009)}]{2009MNRAS.392.1124V}
{Vasudevan}, R.~V., \& {Fabian}, A.~C. 2009, \mnras, 392, 1124, \dodoi{10.1111/j.1365-2966.2008.14108.x}

\bibitem[{Virtanen {et~al.}(2020)Virtanen, Gommers, Oliphant, Haberland, Reddy, Cournapeau, Burovski, Peterson, Weckesser, Bright, {van der Walt}, Brett, Wilson, Millman, Mayorov, Nelson, Jones, Kern, Larson, Carey, Polat, Feng, Moore, {VanderPlas}, Laxalde, Perktold, Cimrman, Henriksen, Quintero, Harris, Archibald, Ribeiro, Pedregosa, {van Mulbregt}, \& {SciPy 1.0 Contributors}}]{2020SciPy-NMeth}
Virtanen, P., Gommers, R., Oliphant, T.~E., {et~al.} 2020, Nature Methods, 17, 261, \dodoi{10.1038/s41592-019-0686-2}

\bibitem[{{Wilkes} {et~al.}(1994){Wilkes}, {Tananbaum}, {Worrall}, {Avni}, {Oey}, \& {Flanagan}}]{1994ApJS...92...53W}
{Wilkes}, B.~J., {Tananbaum}, H., {Worrall}, D.~M., {et~al.} 1994, \apjs, 92, 53, \dodoi{10.1086/191959}

\bibitem[{{Yuan} \& {Narayan}(2014)}]{2014ARA&A..52..529Y}
{Yuan}, F., \& {Narayan}, R. 2014, \araa, 52, 529, \dodoi{10.1146/annurev-astro-082812-141003}

\end{thebibliography}

\end{document}